\documentclass[11pt,a4paper]{article}
\usepackage[english]{babel}
\usepackage[utf8]{inputenc}

\usepackage{slashed}
\usepackage{epsf}
\usepackage{cite}
\usepackage{graphicx}
\usepackage{subcaption} 
\usepackage{verbatim} 
\usepackage{amsmath}
\usepackage{amssymb}
\usepackage{mathrsfs}
\usepackage{mathbbol}
\usepackage{amsfonts}
\usepackage{bbding}
\usepackage{array}
\usepackage[
colorlinks=true
,urlcolor=black
,anchorcolor=black
,citecolor=black
,filecolor=black
,linkcolor=black
,menucolor=black
,linktocpage=true
,pdfproducer=medialab
,pdfa=true
]{hyperref}
\usepackage{color} 

\usepackage[left=2cm,right=2cm,top=2cm,bottom=2cm]{geometry}
\usepackage{tikz}
\usetikzlibrary{shapes,arrows,positioning,automata,backgrounds,calc,er,patterns}
\usepackage[compat=1.1.0]{tikz-feynman}

\vspace {1cm}
\allowdisplaybreaks
\begin{document}
\begin{center}
\vspace*{1mm}
\vspace{1.3cm}
{\Large\bf 
Enhanced di-Higgs production from TeV-scale heavy neutral leptons}

\vspace*{3mm}
{\Large \bf at future lepton colliders}

\vspace*{1.2cm}

{\bf J.~Kriewald $^\text{a}$, E.~Pinsard $^\text{b}$ and A.~M.~Teixeira $^\text{c}$ }

\vspace*{.5cm}
$^\text{a}$ {Jožef Stefan Institut, Jamova Cesta 39, P. O. Box 3000, 1001 Ljubljana, Slovenia}

\vspace*{.2cm}
$^\text{b}$ {Physik-Institut, Universität Zürich, CH-8057 Zürich, Switzerland}

\vspace*{.2cm}
$^\text{c}$ Laboratoire de Physique de Clermont Auvergne (UMR 6533), CNRS/IN2P3,\\
Univ. Clermont Auvergne, 4 Av. Blaise Pascal, 63178 Aubi\`ere Cedex,
France

\end{center}

\vspace*{5mm}
\begin{abstract}
\noindent
Within the context of heavy neutral lepton extensions of the Standard Model, we consider the rare di-Higgs production mode $\ell^+\ell^-\to hh$ at future high-energy lepton colliders.
As a concrete example, we study the impact of a low-scale Inverse Seesaw realisation on the prospects for di-Higgs production. Our results show that the presence of TeV-scale heavy neutral leptons can enhance the cross-section by up to factor 60.
We further comment on the interplay with electroweak precision observables, showing that bounds on the di-Higgs production cross-section at future high-energy lepton colliders could serve as complementary probes of low-scale seesaw scenarios.
\end{abstract}

\section{Introduction}
In recent years, a growing interest has been devoted to future lepton colliders, whose primary aims are to thoroughly explore the scalar sector of the Standard Model (SM), and allow the determination of fundamental Higgs 
parameters and associated observables. 
Likewise, and in view of the nature of the colliding beams, future lepton colliders are also privileged laboratories to carry out precision tests of electroweak (EW) interactions. Among the most promising projects, one can mention future Higgs factories like FCC-ee~\cite{FCC:2025lpp} or CEPC~\cite{CEPCStudyGroup:2023quu},  linear colliders like CLIC~\cite{Brunner:2022usy,Adli:2025swq} or the Linear Collider Facility (LCF)~\cite{LinearCollider:2025lya,LinearColliderVision:2025hlt} and, at the energy frontier, Muon Colliders~\cite{MuonCollider:2022xlm,Accettura:2023ked,MuCoL:2025quu} and plasma wake-field accelerators~\cite{2503.20214}.

In addition to shedding light on the Higgs sector and enabling electroweak precision studies with unprecedented $\mathcal O(10^{-5})$ uncertainties, future lepton colliders also offer extensive opportunities to search for New Physics (see e.g.~\cite{Ai:2024nmn,Blondel:2022qqo,CLIC:2018fvx}). 
Extensions of the SM via heavy neutral leptons (HNL) are one of the most minimal and yet appealing paths towards models of New Physics (NP); in addition to being an integral part of numerous mechanisms of neutrino mass generation, HNL are also naturally present in well-motivated ultraviolet complete models of New Physics (including Left-Right models, grand-unified constructions, among many others). 

\begin{figure}[h!]
    \centering
\raisebox{-6mm}{\begin{tikzpicture}
    \begin{feynman}
    \vertex (ed) at (-0.5, -0.9){\(\ell^+\)};
    \vertex (eu) at (-0.5, 0.9){\(\ell^-\)};
    \vertex (au) at (0.9, 0.9);
    \vertex (bu) at (2.6, 0.9);
    \vertex (ad) at (0.9, -0.9);
    \vertex (bd) at (2.6, -0.9);
    \vertex (hu) at (4,0.9){\( h\)};
    \vertex (hd) at (4,-0.9){\(h\)};
    \diagram  {
    (eu) -- [fermion] (au) -- [boson, edge label=\(Z\)] (bu) -- [scalar] (hu),
    (ed) -- [anti fermion] (ad) --  [boson, edge label'=\(Z\)] (bd) -- [scalar] (hd),
    (au) -- [fermion, edge label'=\(\ell\)] (ad),
    (bu) -- [boson, edge label=\(Z\)] (bd),
    };
    \end{feynman}
    \end{tikzpicture}}
    \hspace{10mm}
    \raisebox{-6mm}{\begin{tikzpicture}
    \begin{feynman}
    \vertex (ed) at (-0.5, -0.9){\(\ell^+\)};
    \vertex (eu) at (-0.5, 0.9){\(\ell^-\)};
    \vertex (au) at (0.9, 0.9);
    \vertex (bu) at (2.6, 0.9);
    \vertex (ad) at (0.9, -0.9);
    \vertex (bd) at (2.6, -0.9);
    \vertex (hu) at (4,0.9){\( h\)};
    \vertex (hd) at (4,-0.9){\(h\)};
    \diagram  {
    (eu) -- [fermion] (au) -- [boson, edge label=\(W\)] (bu) -- [scalar] (hu),
    (ed) -- [anti fermion] (ad) --  [boson, edge label'=\(W\)] (bd) -- [scalar] (hd),
    (au) -- [fermion,edge label'=\(\nu_\ell\)] (ad),
    (bu) -- [boson, edge label=\(W\)] (bd),
    };
    \end{feynman}
    \end{tikzpicture}}\hspace{7mm}\raisebox{7mm}{+ crossed legs}
    \caption{Feynman diagrams of the Standard Model one-loop contribution to $\ell^+\ell^- \to hh$ (displayed in Unitary gauge).}
    \label{fig:SM:eeHH}
\end{figure}
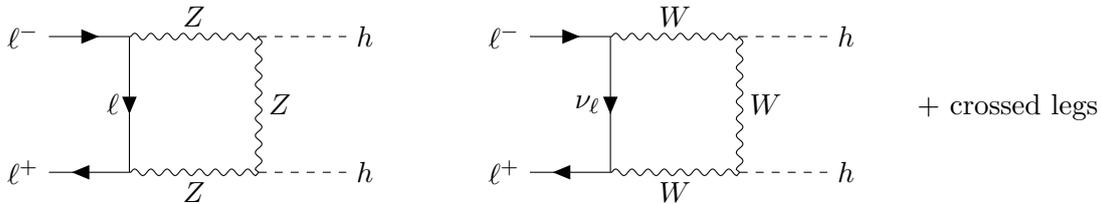

\begin{figure}
    \centering
    \includegraphics[width=0.5\linewidth]{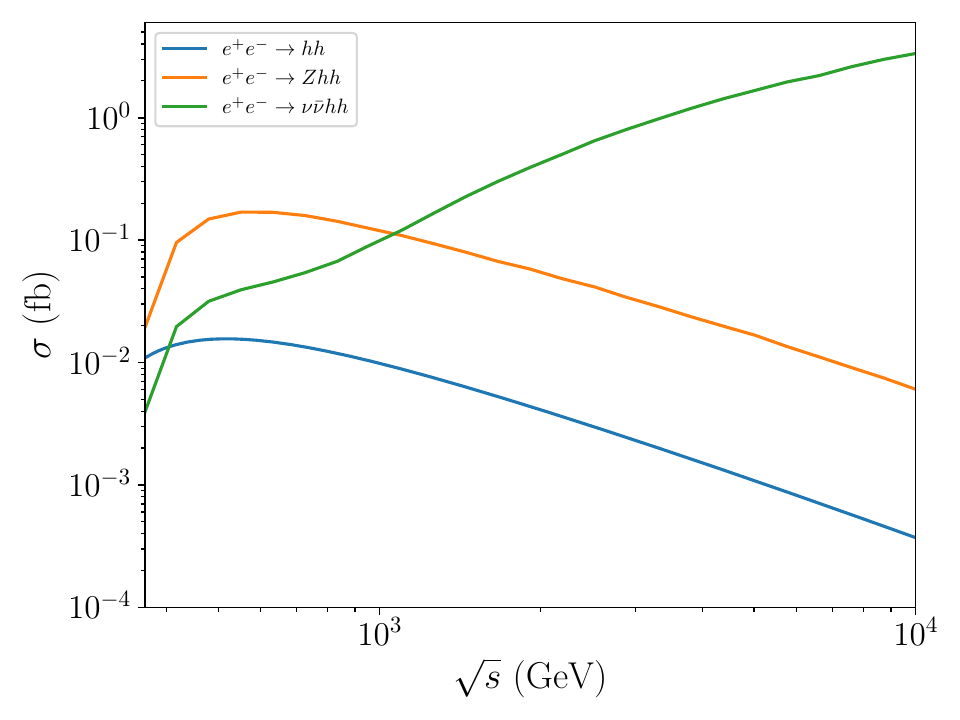}
    \caption{Cross-section predictions for several SM di-Higgs production modes at lepton colliders with respect to the centre-of-mass energy, $\sqrt{s}$.}
    \label{fig:SM_xsecs}
\end{figure}

\begin{figure}
    \centering
    \includegraphics[width=0.48\linewidth]{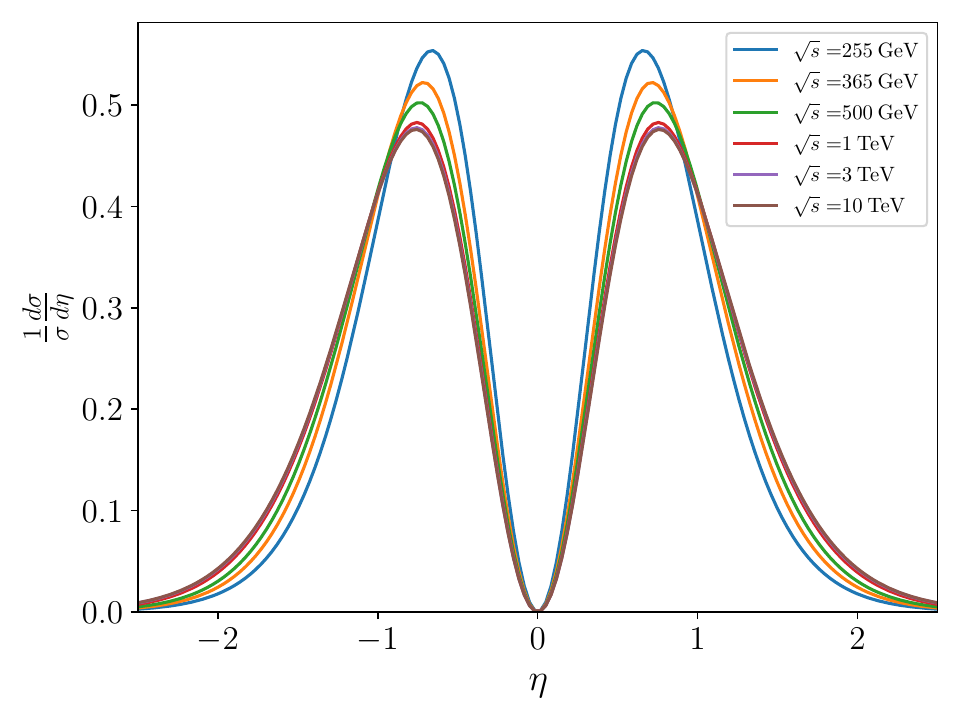}
    \hspace*{4mm}
    \includegraphics[width=0.48\linewidth]{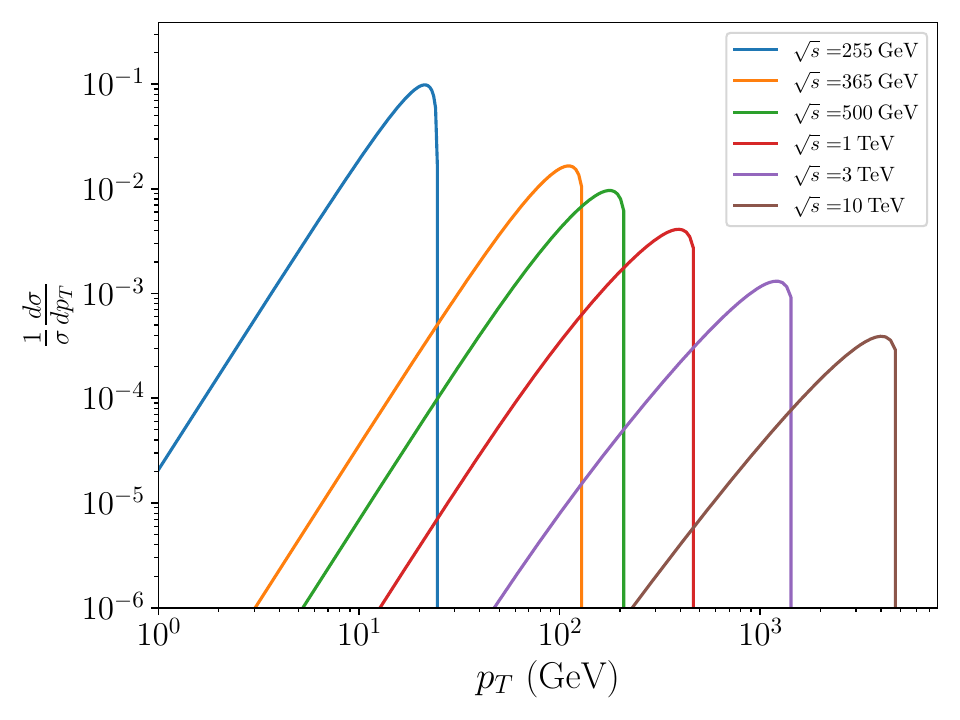}
    \caption{On the left, pseudo-rapidity ($\eta$) distribution of $h$ in $e^+e^-\to hh$ for various fixed $\sqrt{s}$; on the right panel, transverse-momentum ($p_T$) distribution of $h$ in $e^+e^-\to hh$. We consider different possibilities for $\sqrt s$.
    The differential cross-sections have been normalised to their total integrated values.}
    \label{fig:dists}
\end{figure}

As we proceed to discuss, di-Higgs production at lepton colliders offers interesting opportunities to search for NP. 
At future electron-positron colliders, di-Higgs production can occur via several channels: ``double vector boson fusion (VBF)'' $e^+e^-\to W^+W^-\to h h\nu\nu$ and ``double Higgs-strahlung'' $e^+e^-\to Z hh$. 
The (mostly) loop-induced\footnote{The tree-level contributions are of the order $m_e^4/v^4$ and therefore negligible.} process $e^+e^-\to hh$ is not only suppressed due to its higher order (when compared to the tree-level VBF and Higgs-strahlung), but it further suffers from having the $\gamma^*,Z^*\to hh$ amplitudes vanish at one-loop (due to the absence of CP violation in electroweak gauge-scalar interactions, even at the one-loop level).
Further triangle diagrams are proportional to $m_e/v$ (with $v$ the vacuum expectation value of the SM Higgs) and thus vanish in the limit of $m_e\to0$,  being otherwise highly suppressed.
Thus, in the SM, the only relevant contributions to the process $e^+e^-\to hh$ are the box diagrams depicted in Fig.~\ref{fig:SM:eeHH} (see also~\cite{Gaemers:1984vw}).
Moreover, and as can be seen in Fig.~\ref{fig:SM_xsecs}, the SM prediction for the loop-induced $e^+e^-\to h h$ is always sub-dominant compared to other processes such as the ``double Higgs-strahlung'' $e^+e^-\to Z hh $ and the inclusive production channel $e^+e^-\to \nu\bar\nu h h$ (including the double Higgs-strahlung contributions with $Z\to\nu\bar\nu$ and VBF contributions)\footnote{
While the numerical results for the $e^+e^-\to hh$ cross-section rely on the analytical results presented in this paper, the cross-sections for $e^+e^-\to Zhh$ and $e^+e^-\to\nu\bar\nu hh$ have been computed with \textsc{MadGraph5\_aMC@NLO}~\cite{Alwall:2014bza}.}.
The $e^+e^-\to h h$ cross-section has its maximum at $\sqrt{s}\simeq 500$~GeV, reaching about $15$~ab.
This is consistent with previous computations of the SM cross-section found in the literature, see e.g.~\cite{Lopez-Villarejo:2008qii} correcting earlier mistakes in~\cite{Djouadi:1996hp}.
At the $t\bar t$-threshold run with $\sqrt{s}\simeq 360$~GeV the cross-section reaches around $\sim 12$~ab and thus, with optimistic reconstruction efficiencies and the expected integrated luminosity of $2-3\:\mathrm{ab}^{-1}$, one could expect to observe around $10-20$ events.
Furthermore, the $e^+e^-\to h h$ process  has distinctive pseudo-rapidity ($\eta$) and transverse-momentum ($p_T$) distributions, which are shown in Fig.~\ref{fig:dists}.
Together with the absence of forward electrons and/or large missing energy, it should in principle be possible to distinguish the $e^+e^-\to hh$ process from the competing ones, despite its comparatively small cross-section.

In view of the above discussion, it is certainly interesting to consider NP extensions of the SM that explicitly violate CP in the $\gamma,Z\to hh$ interactions, or which can contribute at the loop-level via box diagrams, thus leading to a strong enhancement of the $e^+e^-\to hh$ production cross-section (see e.g.~\cite{Yue:2010zzb} for related discussions in type III seesaw and~\cite{1901.05979} from an effective field theory point of view).
Interestingly, beyond the SM (BSM) constructions including HNL can contribute to these processes:  new box contributions can be present, and the existence of potentially CP violating $Z\,N_i\, N_j$ and $h\, N_i\,N_j$ interactions 
can in principle allow for CP violation in the $Z^*\to hh$ amplitudes. 
However, the latter vertices are (doubly) suppressed; thus only HNL-mediated box diagrams do contribute to $e^+e^-\to hh$ production. 

In the present work, we thus study $e^+e^-\to hh$ production (and $\mu^+\mu^-\to hh$) for SM extensions via HNL.
We compute all relevant higher-order (one-loop) contributions involving the heavy sterile fermions, and provide detailed analytical results. 
As an illustrative representation of a NP realisation including HNL, 
we have considered an appealing mechanism of neutrino mass generation, which can be realised in a natural way at low-energy scales: the Inverse Seesaw (ISS)~\cite{Schechter:1980gr, Gronau:1984ct, Mohapatra:1986bd}. In particular, we focus on an extension of the SM relying on the addition of two species of sterile fermions, $X$ and $\nu_R$ (3 generations of each) - the so-called ISS(3,3).
As extensively discussed in the literature, 
the ISS can have a very rich phenomenology, potentially at the source of contributions to low-energy (flavour) observables~\cite{Abada:2014vea,Abada:2014nwa,Abada:2014kba,Abada:2014cca,Arganda:2015naa,Abada:2015oba,Abada:2018qok,Abada:2024hpb,Kriewald:2024rlg}, (precision) observables at the $Z$-pole~\cite{Arganda:2014dta,DeRomeri:2016gum,2207.10109,2307.02558}, and finally peculiar signatures at colliders~\cite{Das:2012ze,Arganda:2015ija,Das:2016hof,Cai:2017mow,Pascoli:2018heg,Abada:2022wvh,Kriewald:2024cnt}.

In the context of the ISS(3,3), we then carry out a numerical assessment of the prospects regarding the NP contributions to $\ell^+\ell^-\to hh$ (with $\ell = e, \mu$); in order to efficiently survey a multi-dimensional and highly non-trivial parameter space, we rely on new dedicated parametrisations~\cite{Kriewald:2024rlg}, which allow to efficiently encode low-energy data into the Yukawa couplings and other new sources of flavour and CP violation. 
We consider the prospects for several future facilities: in particular, the FCC-ee (running at the $t\bar t$ threshold), a Linear Collider facility (at  $\sqrt{s}=1$~TeV and 3~TeV), a hypothetical plasma wake-field accelerator at 10~TeV and finally a future Muon Collider (at $\sqrt{s}=3$~TeV and 10~TeV). 
Our study suggests that 
significant enhancements to the production cross-section are indeed possible, especially in the case of large centre-of-mass energies. While the FCC-ee does not offer promising prospects, an LCF, or a Muon Collider
will clearly  be sensitive to new contributions to $\ell^+\ell^-\to hh$ arising from the presence of HNL (with typical mass scales between 1~TeV and 10~TeV). 

The $\ell^+\ell^- \to hh $ cross-sections can be increased by the presence of heavy sterile states by up to a factor $60$ for an electron collider running at $\sqrt{s}=10$ TeV, while for the same centre-of-mass energy at a muon collider one can expect an enhancement with respect to the SM prediction of a factor 2.7 (in both cases for HNL masses around $\sqrt{s}/2$). Concerning a more realistic 3 TeV electron collider, the HNL contributions to $\sigma(e^+e^- \to hh) $ can lead to cross-sections as large as twice the expectation of the SM, for heavy state masses around 1.5 TeV. 
For other centre-of-mass energies (and heavy sterile masses), 
one expects modest enhancements, typically up to 10\%. 

As we will emphasise in our discussion, 
it is also important to highlight that searches for HNL via di-Higgs production at future lepton colliders allow searches for such NP models even in the absence of contributions to observables which are usually among the most sensitive probes to the ISS - 
charged lepton flavour violation (cLFV) and electroweak precision observables  (EWPO) (including the invisible $Z$ width).

The manuscript is organised as follows: in Section~\ref{sec:diHiggs} we present the leading HNL contributions to the $\ell^+\ell^-\to hh$ production cross-section, as well as some event shapes in the form of differential cross-sections.
Our most important numerical results are presented in Section~\ref{sec:results}, and we summarise our conclusions in Section~\ref{sec:concs}. The appendices offer complementary information on the considered NP model, and also detailed analytical expressions for the cross-sections.

\section{Di-Higgs production at lepton colliders}
\label{sec:diHiggs}
\begin{figure}[h!]
    \centering
        \raisebox{2mm}{\begin{tikzpicture}
    \begin{feynman}
    \vertex (ed) at (-0.4, -0.9){\(\ell^+\)};
    \vertex (eu) at (-0.4, 0.9){\(\ell^-\)};
    \vertex (au) at (0.9, 0.9);
    \vertex (ad) at (0.9, -0.9);
    \vertex (b) at (2.2,0);
    \vertex (c) at (3.1,0);
    \vertex (hu) at (4.3,0.9){\( h\)};
    \vertex (hd) at (4.3,-0.9){\(h\)};
    \diagram  {
    (eu) -- [fermion] (au) -- [boson, edge label= \(W\)] (b),
    (ed) -- [anti fermion] (ad) -- [boson, edge label'= \(W\)] (b),
    (b) -- [scalar,edge label=\(h\)] (c),
    (hu) -- [scalar] (c)-- [scalar] (hd),
    (au) -- [edge label'=\(n_i\)] (ad),
    };
    \end{feynman}
    \end{tikzpicture}}
    \hspace{3.mm}
    \raisebox{2mm}{\begin{tikzpicture}
    \begin{feynman}
    \vertex (ed) at (-0.4, -0.9){\(\ell^+\)};
    \vertex (eu) at (-0.4, 0.9){\(\ell^-\)};
    \vertex (au) at (0.9, 0.9);
    \vertex (ad) at (0.9, -0.9);
    \vertex (b) at (2.2,0);
    \vertex (c) at (3.1,0);
    \vertex (hu) at (4.3,0.9){\( h\)};
    \vertex (hd) at (4.3,-0.9){\(h\)};
    \diagram  {
    (eu) -- [fermion] (au) -- [ edge label= \(n_i\)] (b),
    (ed) -- [anti fermion] (ad) -- [ edge label'= \(n_j\)] (b),
    (b) -- [scalar,edge label=\(h\)] (c),
    (hu) -- [scalar] (c)-- [scalar] (hd),
    (au) -- [boson, edge label'=\(W\)] (ad),
    };
    \end{feynman}
    \end{tikzpicture}}\hspace{3.mm}
       \raisebox{2mm}{\begin{tikzpicture}
    \begin{feynman}
    \vertex (ed) at (-0.5, -0.9){\(\ell^+\)};
    \vertex (eu) at (-0.5, 0.9){\(\ell^-\)};
    \vertex (au) at (0.9, 0.9);
    \vertex (ad) at (0.9, -0.9);
    \vertex (b) at (2.6,0);
    \vertex (hu) at (4,0.9){\( h\)};
    \vertex (hd) at (4,-0.9){\(h\)};
    \diagram  {
    (eu) -- [fermion] (au) -- [boson, edge label= \(W\)] (b),
    (ed) -- [anti fermion] (ad) -- [boson, edge label'= \(W\)] (b),
    (hu) -- [scalar] (b)-- [scalar] (hd),
    (au) -- [edge label'=\(n_i\)] (ad),
    };
    \end{feynman}
    \end{tikzpicture}}
    \\
   \raisebox{-6mm}{\begin{tikzpicture}
    \begin{feynman}
    \vertex (e1) at (-3.5, -0.9){\(\ell^+\)};
    \vertex (e2) at (-3.5, 0.9){\(\ell^-\)};
    \vertex (i1) at (-2,0);
    \vertex (i2) at (-0.8,0) ;
    \vertex (a1) at (0.5,0.9);
    \vertex (a2) at (0.5,-0.9);
    \vertex (c) at (2,0.9){\( h\)};
    \vertex (d) at (2,-0.9){\(h\)};
    \diagram  {
    (e1) -- [anti fermion] (i1) -- [anti fermion] (e2),
    (i1) -- [boson, edge label=\(Z\)] (i2),
    (a1) -- [edge label=\(n_i\)] (a2) -- [ edge label=\(n_j\)] (i2) -- [edge label=\(n_k\)] (a1),
    (a1) -- [scalar] (c),
    (a2) -- [scalar] (d),
    };
    \end{feynman}
    \end{tikzpicture}}\\
   \raisebox{-7mm}{\begin{tikzpicture}
    \begin{feynman}
    \vertex (ed) at (-0.5, -0.9){\(\ell^+\)};
    \vertex (eu) at (-0.5, 0.9){\(\ell^-\)};
    \vertex (au) at (0.9, 0.9);
    \vertex (bu) at (2.6, 0.9);
    \vertex (ad) at (0.9, -0.9);
    \vertex (bd) at (2.6, -0.9);
    \vertex (hu) at (4,0.9){\( h\)};
    \vertex (hd) at (4,-0.9){\(h\)};
    \diagram  {
    (eu) -- [fermion] (au) -- [edge label=\(n_k\)] (bu) -- [scalar] (hu),
    (ed) -- [anti fermion] (ad) --  [edge label'=\(n_j\)] (bd) -- [scalar] (hd),
    (au) -- [boson, edge label'=\(W\)] (ad),
    (bu) -- [edge label=\(n_i\)] (bd),
    };
    \end{feynman}
    \end{tikzpicture}}
    \hspace{3mm} 
 \raisebox{-9mm}{\begin{tikzpicture}
    \begin{feynman}
    \vertex (ed) at (-0.5, -1.5){\(\ell^+\)};
    \vertex (eu) at (-0.5, 0.8){\(\ell^-\)};
    \vertex (au) at (0.9, 0.8);
    \vertex (bu) at (2.6, 0.8);
    \vertex (ad) at (0.9, -0.8);
    \vertex (bd) at (2.6, -0.8);
    \vertex (hu) at (4,0.8){\( h\)};
    \vertex (hd) at (4,-1.5){\(h\)};
    \diagram  {
    (eu) -- [fermion] (au) -- [boson, edge label=\(W\)] (bu) -- [scalar] (hu),
    (ed) -- [anti fermion] (bd) --  [ edge label'=\(n_j\)] (ad) -- [scalar] (hd),
    (au) -- [edge label'=\(n_i\)] (ad),
    (bu) -- [boson, edge label=\(W\)] (bd),
    };
    \end{feynman}
    \end{tikzpicture}}
    \hspace{3mm}
\raisebox{-6.5mm}{\begin{tikzpicture}
    \begin{feynman}
    \vertex (ed) at (-0.5, -0.9){\(\ell^+\)};
    \vertex (eu) at (-0.5, 0.9){\(\ell^-\)};
    \vertex (au) at (0.9, 0.9);
    \vertex (bu) at (2.6, 0.9);
    \vertex (ad) at (0.9, -0.9);
    \vertex (bd) at (2.6, -0.9);
    \vertex (hu) at (4,0.9){\( h\)};
    \vertex (hd) at (4,-0.9){\(h\)};
    \diagram  {
    (eu) -- [fermion] (au) -- [boson, edge label=\(W\)] (bu) -- [scalar] (hu),
    (ed) -- [anti fermion] (ad) --  [boson, edge label'=\(W\)] (bd) -- [scalar] (hd),
    (au) -- [edge label'=\(n_i\)] (ad),
    (bu) -- [boson, edge label=\(W\)] (bd),
    };
    \end{feynman}
    \end{tikzpicture}}
    \caption{Feynman diagrams of the one-loop contribution to $\ell^+\ell^- \to hh$ in the presence of HNL (displayed in Unitary gauge).}
    \label{fig:eehh}
\end{figure}
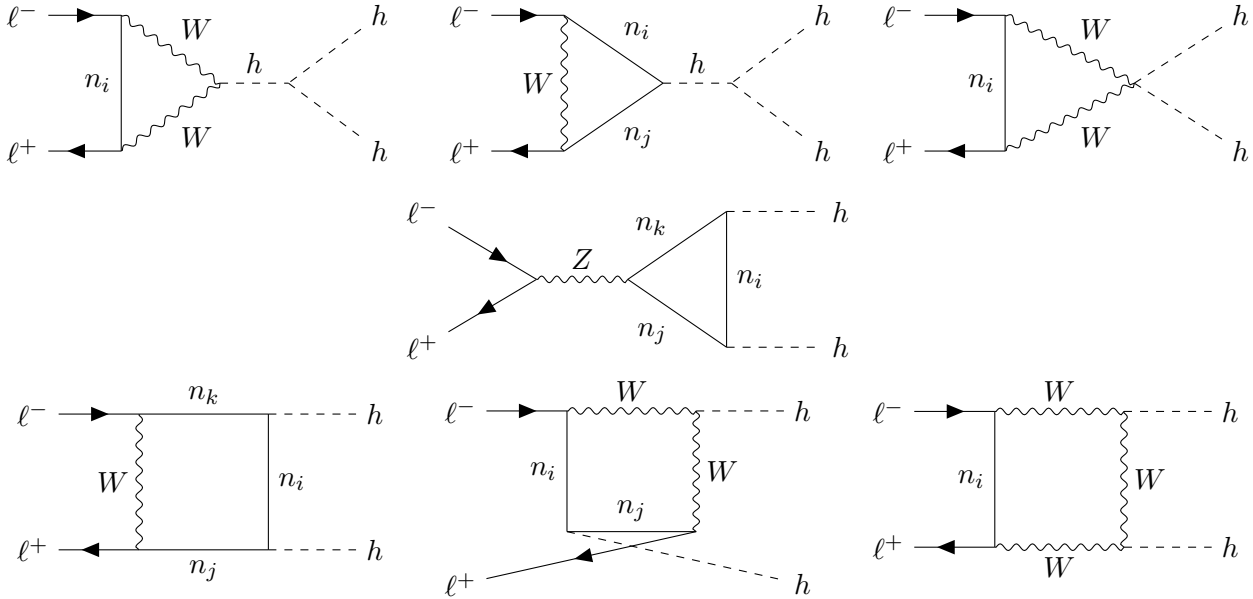

In the presence of additional heavy sterile fermions, there are several new contributions to the loop-induced Higgs pair production cross-section, as displayed in Fig.~\ref{fig:eehh}.
Just as it occurs in the SM, the first class of diagrams (first row of Fig.~\ref{fig:eehh}), in essence corresponds to a contribution to the renormalisation of the electron Yukawa coupling, and therefore vanishes in the limit of $m_e\to 0$.
The second class concerns the triangle diagrams with a virtual $Z^*$ exchange (second row of Fig.~\ref{fig:eehh}).
While the entire diagram topology of $Z^*\to hh$ vanishes in the SM due to CP-conservation, this is not necessarily the case in the presence of HNL exchanges.
Due to their Majorana nature, the effective $Z-N_iN_j$ and $h-N_iN_j$ vertices are flavour and CP violating (see Eq.~\eqref{eqn:lag}). 

The resulting amplitudes from the $Z^*\to hh$ diagrams always have a particular mixing structure,
\begin{equation}
    \mathcal A_{Z^*\to hh} \,\propto \,\sum_{i,j,k}\mathrm{Im}(\mathcal C_{ij}\mathcal C_{jk}\mathcal C_{ki}^*) + \,\text{variations}\,,
\end{equation}
which depends on the flavour and CP violating couplings encoded in $\mathcal C_{ij}$, which is defined as  
\begin{equation}\label{eq:Cij}
    \mathcal C_{ij} \,= \,\sum_\rho \mathcal U_{\rho i}\mathcal U_{\rho j}^*\,,
\end{equation}
with  $\mathcal U_{\alpha i}$ the $3\times n$ rectangular lepton mixing matrix (see Appendix~\ref{app:model}).
As can be seen, the $Z^*\to hh$ amplitudes vanish in the limit of CP-conservation (real $\mathcal C_{ij}$) and/or if identical HNL propagate in the loop.
For three distinct HNL - as in the case of the ISS(3,3) - and for complex $\mathcal C_{ij}$, the $Z^*\to hh$ amplitude can receive non-vanishing contributions, but is nevertheless very small due to mixing suppression: any contribution in $\mathcal C_{ij}$ with $i,j> 3$ scales as $v^2 Y_D^2/m_N^2$ (heavy-light mixing squared), such that three insertions of $\mathcal C_{ij}$ with heavy $n_i, n_j$ lead to six-fold mixing-suppressed contributions.
Also notice that while some of this suppression is alleviated due to the heavy masses appearing in the Feynman rules (see Eq.~\eqref{eqn:lag}) and entering in the loop-functions, the diagrams also scale as $\propto 1/s$, rendering them negligible with respect to the SM contributions at high energies.

The remaining contributions are from the box-diagrams in the third row of Fig.~\ref{fig:eehh} with either one, two or three virtual heavy $n_i$ in the loop.
While we do not display the lengthy formulae for the box contributions, all the relevant expressions (for the SM and HNL contributions) can be found in Appendix~\ref{app:analytical} in terms of Passarino-Veltman functions.
A similar discussion of mixing suppression applies also here with the exception of the boxes containing a single $n_i$ in the loop.
Here, certain terms proportional to $|\mathcal U_{e i}|^2m_i^2$ appear (see Eq.~(\ref{eqn:goldt})). 
Furthermore, the box-contributions do not suffer from a $1/s$ suppression in the high-energy limit such that sizeable enhancements with respect to the SM are possible.

Due to the single external fermion current, the amplitude has a particularly simple helicity decomposition.
After a lengthy but straight-forward evaluation of the box diagrams one finds
\begin{equation}
    \frac{d\sigma}{d\cos\theta} \,=\, \frac{1}{32\pi \,s}\sqrt{1 - 4 \frac{m_h^2}{s}}\, \frac{1}{2}\,\left[\frac{m_h^4 }{4} \,t\, u
    \left[(1 - \lambda_-)(1 + \lambda_+)|F_L|^2 + (1 + \lambda_-)(1+\lambda_+)|F_R|^2\right]\right]\,,
    \label{eqn:dsigdc}
\end{equation}
with $\lambda_\mp$ the polarisation of the electron (positron) beams, and the usual Mandelstam variables
\begin{equation}
    s \,=\, (p_1 + p_2)^2\,,\quad t \,=\, (p_1 - p_3)^2\,,\quad 
    u \,=\, (p_1 - p_4)^2\,,
\end{equation}
in which $p_{1 (2)}$ denote the incoming electron (positron) momentum and $p_{3,4}$ the outgoing Higgs momenta.
The scattering angle is defined between $p_3$ and $p_1$, and can be written in terms of invariants as
\begin{equation}
    \cos\theta \,=\, \left(1 - \frac{4\, m_h^2}{s}\right)^{-1/2}\, \frac{2t + s - 2m_h^2}{s}\,,
\end{equation}
with $\cos\theta\in(-1,1)$.
The Mandelstam variables $t,u$ can be rewritten in terms of $\cos\theta$,
\begin{eqnarray}
    t &=& -\frac{s}{2}\left(1 - 2 \frac{m_h^2}{s} - \sqrt{1-4\frac{m_h^2}{s}}\cos\theta\right)\\
		u &=& -\frac{s}{2}\left(1 - 2 \frac{m_h^2}{s} + \sqrt{1-4\frac{m_h^2}{s}}\cos\theta\right)\,.
\end{eqnarray}
The total cross-section can then be obtained by (numerically) integrating the differential cross-section in Eq.~\eqref{eqn:dsigdc} over $\cos\theta$.
The form-factor $F_{L}$ receives contributions from all types of diagrams, while $F_R$ only arises from the SM diagram containing three virtual $Z$-bosons (see Fig.~\ref{fig:SM:eeHH}).
One can further decompose the contributions from ``straight'' and ``crossed-legged'' boxes (denoted by $t$ and $u$ subscripts), leading to
\begin{equation}
    F_L \, = \, F_{L,t}^Z + F_{L,t}^W + F_{L,t}^{3N} + F_{L,t}^{2N} + F_{L,t}^{N} + t\leftrightarrow u\,\quad\text{and}\quad 
    F_R \,= \,F_{R,t}^Z + F_{R,u}^Z\,,
\end{equation}
with $F_{L(R),t(u)}^{xx}$ as given in Appendix~\ref{app:analytical}.

From the differential cross-section in Eq.~\eqref{eqn:dsigdc} we can further derive the pseudo-rapidity distribution, as well as the transverse momentum distribution, via 
\begin{equation}
    \cos\theta = \pm \sqrt{1 - \frac{p_T^2}{p_h^2}}\,,\quad \text{and}\quad \cos\theta = \tanh \eta\,,
\end{equation}
and the appropriate Jacobians
\begin{equation}
    \left|\frac{d\cos\theta}{dp_T}\right| \,=\, \frac{p_T}{p_h^2\sqrt{1 - (p_T/p_h)^2}}\,,\quad\text{and}\quad\left|\frac{d\cos\theta}{d\eta}\right| \,=\, 1 - \tanh^2\eta\,,
\end{equation}
in which the Higgs momentum $p_h$ is given by $p_h = \frac{1}{2}\sqrt{s - 4m_h^2}$.
We thus obtain for the $p_T$ distribution
\begin{equation}
    \frac{d\sigma}{dp_T} \,=\, 
    \frac{p_T}{p_h^2\sqrt{1 - (p_T/p_h)^2}}\left[\left.\dfrac{d\sigma}{d \cos\theta}\right|_{\cos\theta = + \sqrt{1 - (p_T/p_h)^2}} \,+\, \left.\dfrac{d\sigma}{d \cos\theta}\right|_{\cos\theta = - \sqrt{1 - (p_T/p_h)^2}}\right]\,,
\end{equation}
and for the pseudo-rapidity distribution
\begin{equation}
    \frac{d\sigma}{d\eta} \,=\, (1 - \tanh^2\eta)\left.\frac{d\sigma}{d\cos\theta}\right|_{\cos\theta=\tanh\eta}\,.
\end{equation}
For the Standard Model, the integrated cross-sections, as well as the pseudo-rapidity and $p_T$ distributions, have been already shown in Figs~\ref{fig:SM_xsecs}~and~\ref{fig:dists} (see the Introduction).

Since the dominant HNL contribution - $F_L^N$ - exhibits the same 
kinematical dependence as in the leading SM contributions, 
the relative $p_T$ and $\eta$ distributions remain unchanged (see Fig.~\ref{fig:dists}) but - and as we will subsequently argue - the  total cross-section can receive significant enhancements.
While we do not explicitly study the effects of beam polarisation, we nevertheless notice that polarisations of $P(e^-,e^+)= (-0.8,0.3)$ or $P(e^-,e^+)= (-0.8,0.6)$  (achievable at linear colliders) can further enhance the cross-sections by factors of $\sim2.43$ or $\sim2.88$, respectively.
(The kinematics of the process remain mostly unaffected by beam polarisation, due to the helicity suppression of the $F_R$ form-factor.)

After this brief discussion we proceed to numerically assess the NP contributions, focusing - and as mentioned before - on the Inverse Seesaw, a well-motivated low-scale realisation, that potentially leads to sizeable enhancements of the cross-sections.
We nevertheless emphasise that the analytical results presented here and in the Appendix~\ref{app:analytical} are valid for BSM constructions via heavy sterile states (featuring enlarged leptonic mixings).
\section{Results}\label{sec:results}
In what follows we summarise our most important numerical results concerning the impact of the new sterile states on the di-Higgs production cross-section. 
We also comment on the heavy state contributions to electroweak observables and other flavour-conserving observables. 
As described in Appendix~\ref{app:model}, we consider the ISS(3,3), and relying on the ``ISS-SVD'' parametrisation developed in~\cite{Kriewald:2024rlg}, we focus on realisations generically leading to charged lepton flavour conservation.
Furthermore, we consider mixing scenarios that saturate the model-independent bounds obtained in~\cite{Blennow:2023mqx}.
\subsection{New contributions to $\pmb{\sigma(\ell^+ \ell^- \to hh)}$: electron and muon colliders}
In the present study, our primary goal is to assess the overall impact of the new HNL-mediated contributions of the di-Higgs production cross-sections at future lepton colliders (without aiming at realistic simulations). 
In particular, we focus on three future facilities: the FCC-ee (in its $t \bar t$ run), Linear Collider facilities~\cite{CLICdp:2018cto,LinearCollider:2025lya,LinearColliderVision:2025hlt} and a future Muon Collider~\cite{MuonCollider:2022nsa,MuonCollider:2022xlm,MuCoL:2025quu,Accettura:2023ked}. 
For the last two, we consider illustrative benchmark points for (high) centre-of-mass energies. 
The information is summarised in Table~\ref{tab:sqrts_facilities}, in which we also include 
the expected targets for the integrated luminosities.  
While there is no concrete design for a 10~TeV $e^+e^-$ collider yet, the ongoing development of plasma wake-field accelerators (PWA) could in principle render feasible such a machine~\cite{2503.20214}, with similar luminosity targets of $\mathcal O(1-10)\:\mathrm{ab}^{-1}$.
Even if our goal is not to conduct a complete feasibility study of measuring the $\ell^+\ell^-\to hh$ process, the envisaged luminosities 
should allow setting stringent bounds on the SM cross-section or even discover $\ell^+\ell^-\to hh$ in case of a significant enhancement due to BSM contributions. 
\renewcommand{\arraystretch}{1.3}
\begin{table}[]
    \centering
    \begin{tabular}{|c|c|c|}
    \hline
        Facility & $\sqrt{s}$ & luminosity \\\hline\hline
        FCC-ee~\cite{FCC:2025lpp} & 365 GeV & 2.7 ab$^{-1}$\\ \hline
        LCF~\cite{LinearCollider:2025lya,LinearColliderVision:2025hlt} & 1 TeV & $\mathcal O(5-10)\:\mathrm{ab}^{-1}$\\
        CLIC~\cite{CLICdp:2018cto}& 3 TeV & 5 ab$^{-1}$\\ 
        PWA~\cite{2503.20214} & 10 TeV & $\mathcal O(1-10)\:\mathrm{ab}^{-1}$\\
        \hline
        Muon Collider~\cite{MuCoL:2025quu} & 3 TeV & 1 ab$^{-1}$\\
        & 10 TeV & 10 ab$^{-1}$\\ \hline
    \end{tabular}
    \caption{Expected centre-of-mass energies and integrated luminosities for different future lepton colliders: FCC-ee, LCF, CLIC and Muon Collider. }
    \label{tab:sqrts_facilities}
\end{table}

\begin{figure}[h!]
    \centering
    \includegraphics[width=0.7\linewidth]{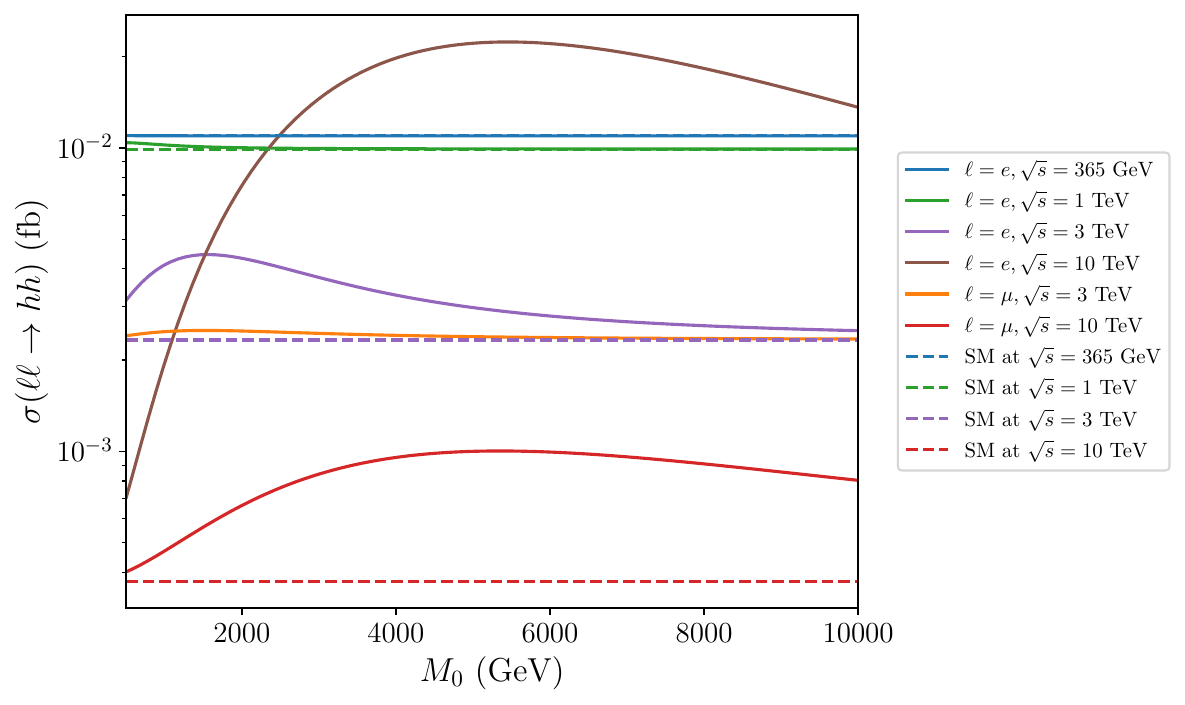}
    \caption{Cross-section for $\ell \ell \to hh$ production
    (fb) as a  function of the degenerate heavy neutrino mass ($M_0$, in GeV), for $\sqrt{s}=365$~GeV, 1~TeV, 3~TeV and 10~TeV, and colliding electron or muon beams. Full (dashed) lines correspond to the total (SM) contributions to the di-Higgs production cross-section.}
    \label{fig:sig_vs_mass}
\end{figure}

In Fig.~\ref{fig:sig_vs_mass} we begin by considering the 
$\ell^+ \ell^- \to hh$ production cross-sections versus the mass of the (degenerate) heavy sterile spectrum\footnote{For simplicity, in what follows we will consider $M_R^{ij}=M_0\, \delta_{ij}$, see Eq.~(\ref{eq:ISS:lagrangian}).}, denoted $M_0$. We display the full contributions and the SM ones for distinct centre-of-mass energies, as summarised in Table~\ref{tab:sqrts_facilities}.
As can be seen, for $e^+e^-$ colliders operating at $\sqrt{s}\leq 1$~TeV, there is only a small $\%$-level enhancement with respect to the SM prediction for up to TeV-scale $M_0$.
At higher centre-of-mass energies, as would be the case of PWAs or muon colliders, the cross-section can be significantly enhanced, reaching its peak for $M_0\simeq \sqrt{s}/2$.
The most striking and experimentally promising prospects emerge for a PWA $e^+e^-$ machine operating at $\sqrt{s}=10$~TeV: in this case, the NP contributions lead to an enhancement of the production cross-section that can reach as much as a factor 60, up to 
$20$~ab.
The different behaviour of the $e^+e^-\to hh$ and $\mu^+\mu^-\to hh$ cross-sections is due to different indirect bounds on the involved heavy-light mixings, as encoded in $\eta_{ij}$ (see Appendix~\ref{app:model}).
Since the bound on muon mixings constraining $\eta_{\mu\mu}$ is 10 time stronger than the associated one on $\eta_{ee}$, 
the enhancements of the $\mu^+\mu^-\to hh$ cross-section are thus smaller, but can still lead to enhancements up to around a factor $\mathcal O(3)$.

\begin{figure}[h!]
    \centering
    \includegraphics[width=0.5\linewidth]{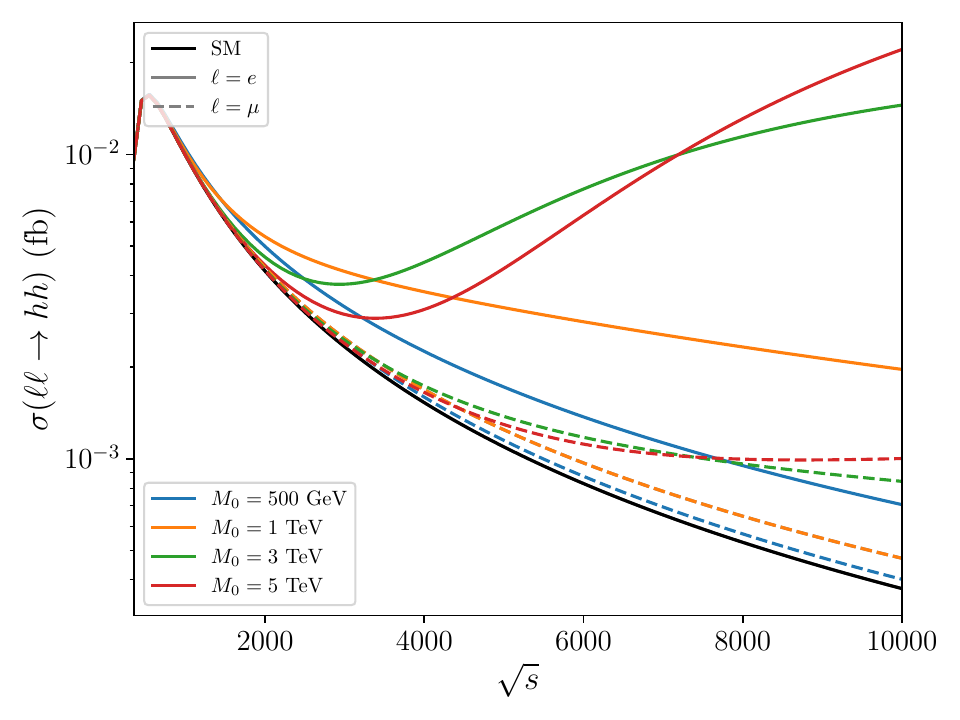}
    \caption{Cross-section for $\ell \ell \to hh$ production
    (fb) as a  function of the centre-of-mass energy $\sqrt{s}$, for different degenerate heavy neutrino masses, $M_0=0.5$~TeV, 1~TeV, 3~TeV and 5~TeV; the solid lines correspond to $ee \to hh$ and the dashed ones to $\mu \mu \to hh$. We further display the (flavour-universal) SM prediction.}
    \label{fig:sig_vs_sqrts}
\end{figure}

A complementary view is offered in Fig.~\ref{fig:sig_vs_sqrts}, in which the $\ell^+ \ell^- \to hh$ production cross-sections are now presented as a function of the centre-of-mass energy $\sqrt{s}$, for illustrative regimes of the (degenerate) heavy neutrino spectrum.

\begin{figure}[h!]
\centering
\includegraphics[width=0.48 \textwidth]{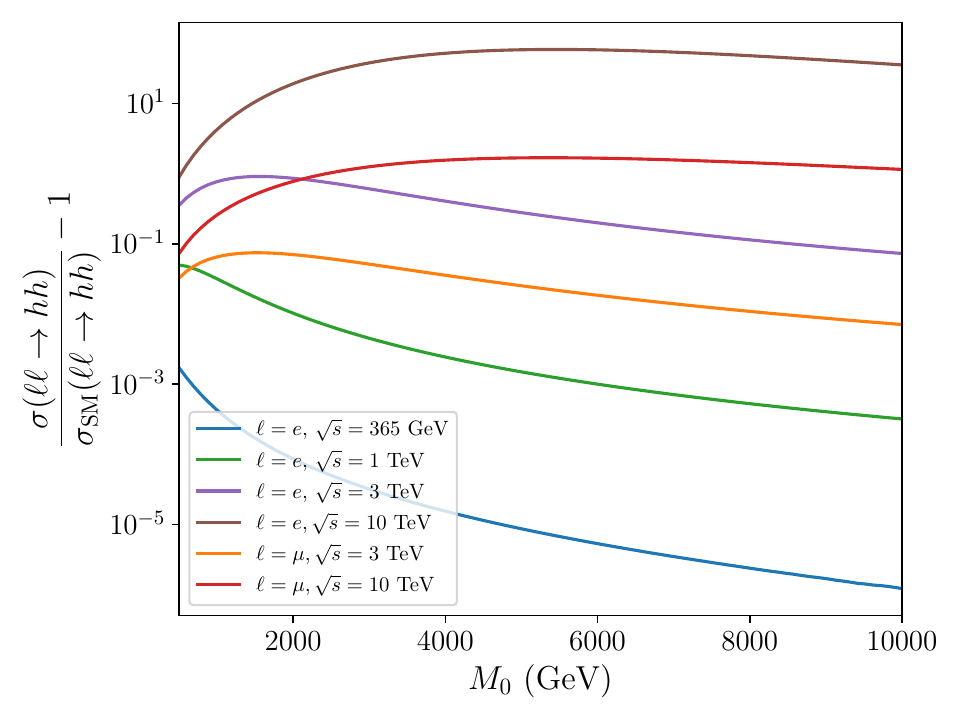}
\hspace*{4mm}
\includegraphics[width=0.48 \textwidth]{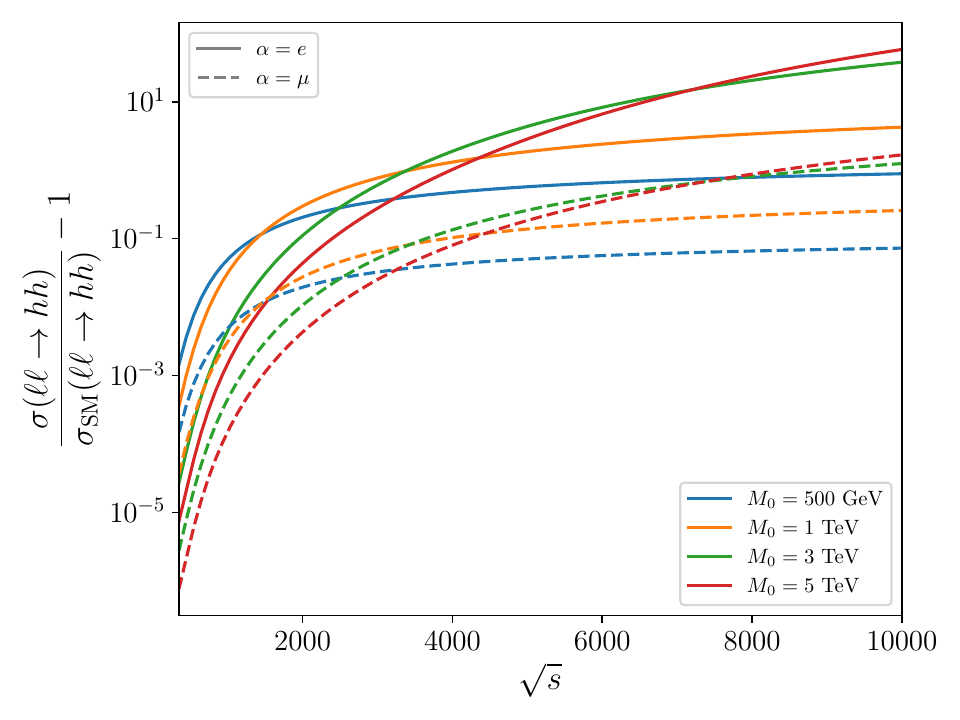} 
\caption{Relative contributions with respect to the SM cross-section for $\ell \ell \to hh$: on the left, vs. the degenerate heavy sterile masses (and for the considered $\sqrt s$ regimes); on the right, as a function of $\sqrt s$, and for several values of the heavy neutrinos.}
    \label{fig:deltavsMs}
\end{figure}

For further clarity, in Fig.~\ref{fig:deltavsMs}, we now display the relative enhancement of the cross-section, versus the heavy sterile masses and the centre-of-mass energy for the considered lepton colliders.
The results shown in Figs.~\ref{fig:sig_vs_sqrts}~and~\ref{fig:deltavsMs} further underline that TeV-scale HNL contributions can lead to sizeable enhancements of factor $\mathcal O(10-60)$. 
In order to summarise our findings, the maximum enhancements of the cross-sections, for some benchmark choices of the HNL masses, are collected in Table~\ref{tab:enhanceSM}.
\begin{table}[h!]
    \centering
    \begin{tabular}{|c|c|c|c|c|}
    \hline 
     & \multicolumn{4}{c|}{Values of $\sigma(\ell\ell \to hh) $/$\sigma_{\mathrm{SM}}(\ell\ell \to hh) $} \\
    \hline
       HNL masses & $ee$ @ 3 TeV  & $\mu\mu$ @ 3 TeV  & $ee$ @ 10 TeV & $\mu\mu$ @ 10 TeV \\ \hline \hline
        500 GeV & 1.4 & 1.03 & 1.9 & 1.07 \\
        1 TeV & 1.8 & 1.06 & 2.7 & 1.25 \\
        3 TeV & 1.6 & 1.05 & 40 & 2.2 \\
        5 TeV & 1.3 & 1.03 & 60 & 2.75 \\ \hline
    \end{tabular}
    \caption{Enhancement of the cross-sections with respect to the SM expectations, for electron and muon colliders, at $\sqrt{s}=3$~TeV and 10 TeV, for the considered degenerate HNL masses.}
    \label{tab:enhanceSM}
\end{table}

\subsection{Interplay with electroweak precision observables}
In our numerical analysis, we have explored regions of ISS parameter space which are explicitly flavour conserving (up to neutrino oscillation data).

As we have shown, sizeable enhancements of the $\ell^+\ell^-\to hh$ cross-sections are possible, opening a complementary window to probe this class of models at high-energy lepton colliders. As we have discussed, 
the enhancements are driven by sizeable active-sterile mixings, which we have taken close to values saturating the model-independent low-energy bounds~\cite{Blennow:2023mqx}.
As pointed out in~\cite{2307.02558}, such  
large mixings can further lead to important contributions to electroweak precision and lepton flavour conserving observables.
Thus, we also address here the impact of the considered ISS scenario on the flavour conserving Higgs and $Z$ leptonic decays, $h,Z\to\ell^+\ell^-$ decays (at one-loop level), 
on the loop-corrected $Z\to$~invisible width, as well as on the oblique  $T$-parameter.
The analytical results which we base our present analysis on can be found in~\cite{2307.02558}.

\begin{figure}
    \centering
    \includegraphics[width=0.48\linewidth]{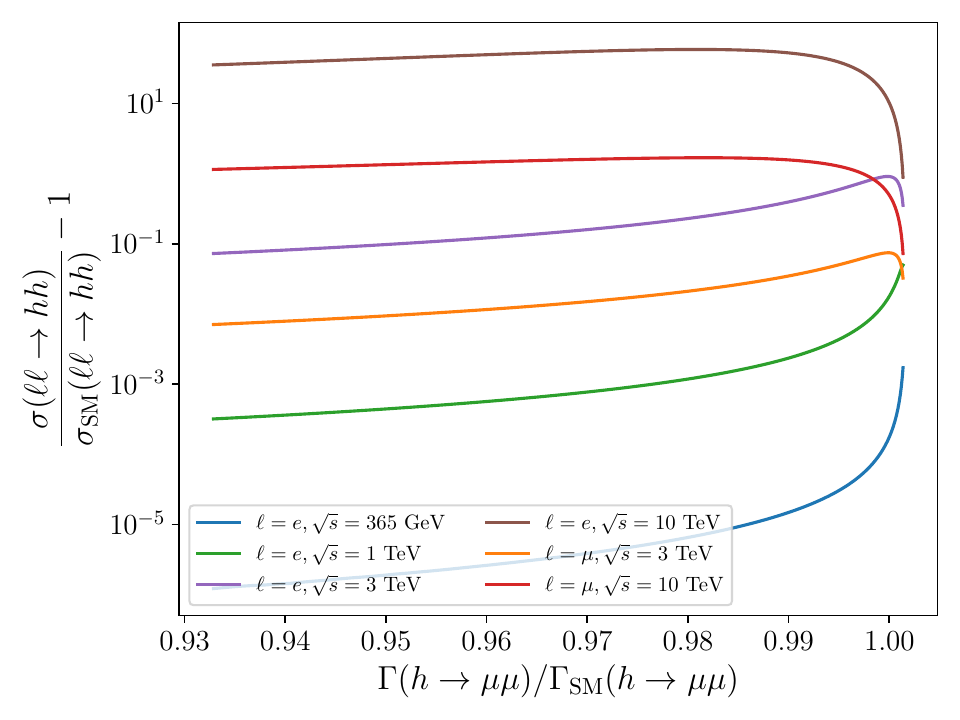}
    \hspace*{4mm}
    \includegraphics[width=0.48\linewidth]{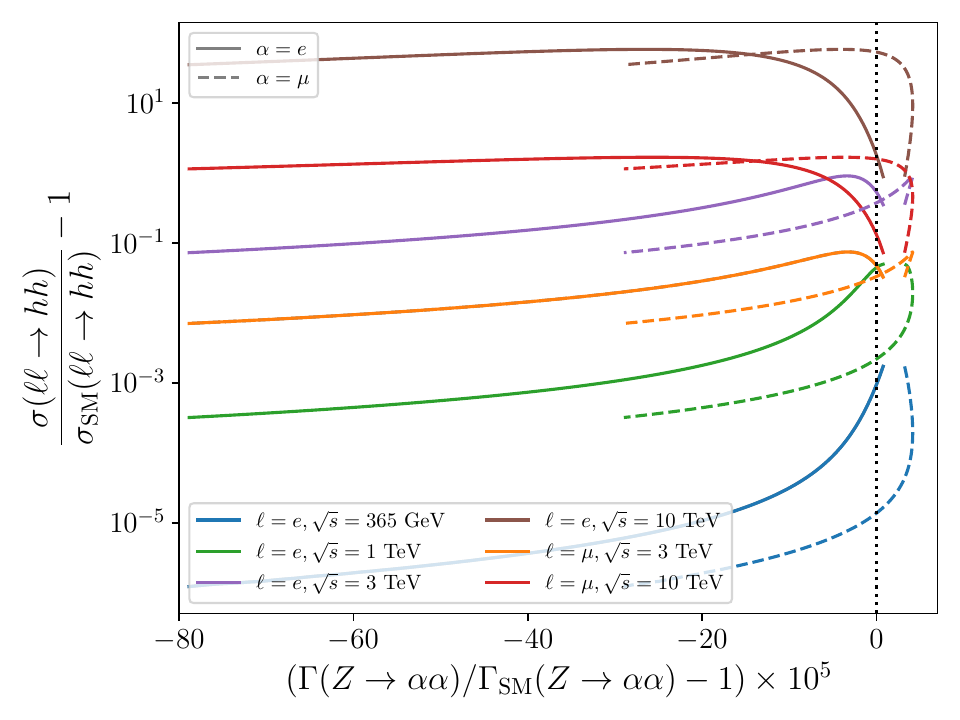}
    \caption{On the left, relative deviations of $\sigma(\ell \ell \to hh)$ with respect to the SM production cross-section vs. the $H\to \mu\mu$ width (normalised to the SM prediction), for the considered $\sqrt s$ regimes. On the right, relative deviations of $\sigma(\ell \ell \to hh)$ as a function of the relative deviations in $Z\to \alpha\alpha$ for $\alpha=e$ (solid lines) and $\alpha=\mu$ (dashed lines), again for the considered $\sqrt s$ regimes.}
    \label{fig:flavour}
\end{figure}

As in the previous sub-section, we fix the heavy-light mixings to saturate the model-independent bounds on the ``non-unitarity'' parameters $\eta_{ij}$ (see Appendix~\ref{app:constraints}) and consider a degenerate heavy spectrum with $M_R^{ij} = M_0\,\delta_{ij}$.
In Fig.~\ref{fig:flavour} we confront the relative enhancement of the $\ell^+\ell^-\to hh$ cross-sections with the associated predictions for the loop-corrected $h\to\mu\mu$ and $Z\to\ell\ell$ widths (normalised to the SM values).
As can be seen, large enhancements of the $\ell^+\ell^-\to hh$ cross-sections can also lead to an $\mathcal O(5\%-10\,\%)$ suppression of the $h\to\mu\mu$ rates, due to the destructive interference of one-loop HNL contributions  with the SM electroweak amplitudes (at next to leading order (NLO)).
A similar effect, albeit much milder, is visible for $Z\to ee,\,\mu\mu$ decays. 
Here, the interference of the HNL contributions leads to a suppression of the decay widths, $\mathcal O(10^{-4}-10^{-3})$, with respect to the SM prediction at next-to-NLO (NNLO)~\cite{Freitas:2014hra}.
While this suppression is comparatively much smaller, future measurements at the FCC-ee $Z$-pole run are anticipated to reach an unprecedented relative precision of $\mathcal O(10^{-5}-10^{-4})$, thus rendering testable very small departures from the SM prediction.

It is important to stress that sizeable enhancements of the $\ell^+\ell^-\to hh$ cross-sections can occur even in the absence of significant deviations from the SM expectation of  $h\to\mu\mu$ and $Z\to\ell\ell$ decays. This clearly highlights the complementarity role of leptonic di-Higgs production with respect to the above mentioned observables.

\begin{figure}
    \centering
    \includegraphics[width=0.48\linewidth]{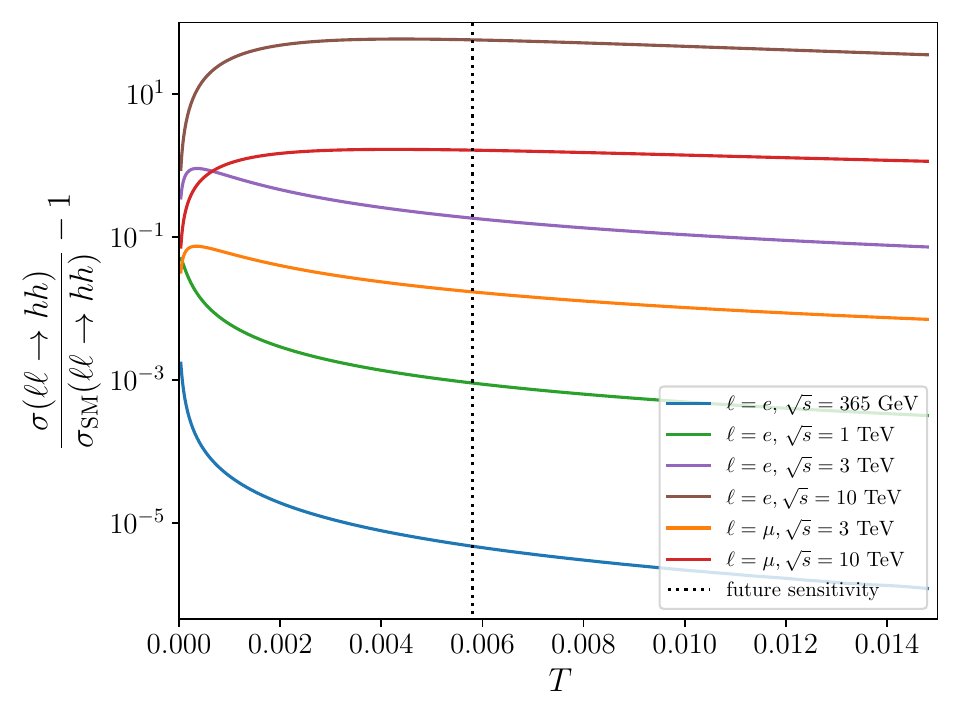}
    \hspace*{4mm}
     \includegraphics[width=0.48\linewidth]{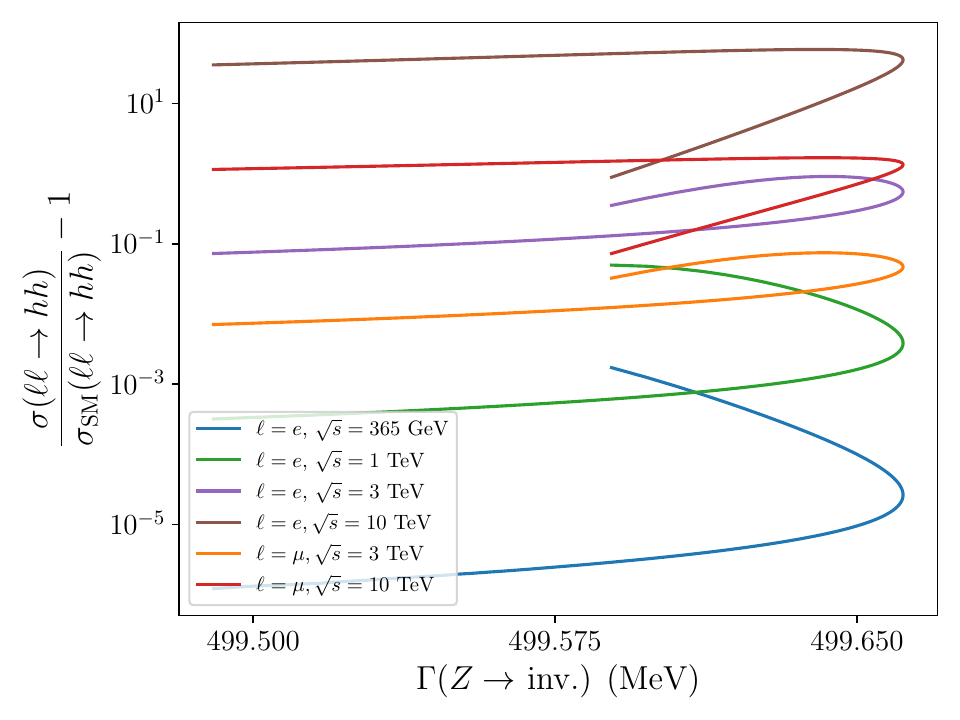}
    \caption{Relative deviations of $\ell \ell \to hh$ with respect to the SM cross-section versus the $T$-parameter (left panel) and versus  the loop-corrected rate $\Gamma(Z\to\text{inv})$. 
    The dotted line on the left panel represents the future FCC-ee sensitivity. }
    \label{fig:noflavour}
\end{figure}

\bigskip
It is also interesting to consider the EW observables,  $\Gamma(Z\to\text{inv})$ (at one-loop) and the $T$-parameter.
In Fig.~\ref{fig:noflavour} thus we present the relative deviations of the $\ell^+\ell^-\to hh$ cross-sections versus the  $T$-parameter and the prospects for $\Gamma(Z\to\text{inv})$.
The results for the $T$-parameter strengthen our previous conclusions: sizeable enhancements of the di-Higgs cross-section can lead, but do not necessarily imply, measurable effects in the $T$-parameter. 
As can be seen from the left panel of Fig.~\ref{fig:noflavour}, 
the $T$-parameter ranges up to $\sim 0.015$: while such values remain clearly  below the current exclusion limits ($T=0.047\pm 0.066$~\cite{ParticleDataGroup:2024cfk}), they will be within reach of FCC-ee, which is expected to be sensitive to $T\lesssim 0.0058$~\cite{deBlas:2019rxi}.

Contrary to the $T$-parameter and the flavour dependent $h\to\mu\mu$ and $Z\to\ell\ell$ rates, the effect for $\Gamma(Z\to\text{inv})$ is much stronger.
The SM prediction for the invisible $Z$ decay rate is $\Gamma_\text{SM}(Z\to\text{inv})\simeq501.45\pm0.05$~MeV~\cite{Freitas:2014hra}, while fits to LEP and LHC data indirectly determine $\Gamma_\text{exp}(Z\to\text{inv})\simeq499.3 \pm 1.5$~MeV~\cite{ParticleDataGroup:2024cfk}.
As can be seen on the right panel of Fig.~\ref{fig:noflavour}, the regimes in the considered ISS parameter space leading to sizeable enhancements of the leptonic di-Higgs production cross-section are associated to predictions of $\Gamma_\text{ISS}(Z\to\text{inv})\simeq 499.5-499.7$~MeV.
The precision of future measurements at FCC-ee should improve the uncertainties on $\Gamma(Z\to\text{inv})$ by at least an order of magnitude~\cite{FCC:2018byv,FCC:2018evy}, rendering this scenario directly testable.

\section{Conclusions and Outlook}\label{sec:concs}
In our work we have studied the impact of new physics contributions regarding di-Higgs production at future lepton colliders, 
$\sigma(\ell^+ \ell^- \to hh)$. In particular, we have considered 
SM extensions via HNL, focusing on a complete model of neutrino mass generation, the Inverse Seesaw. While we relied on the ISS(3,3) as a canvas for our numerical studies, we emphasise that the expressions here derived for the production cross-section are generically valid for low-scale fermion singlet seesaw realisations. 
 
As we have discussed here, in the ISS(3,3) the contributions of the
heavy sterile fermions (box diagrams) can lead to sizeable enhancements of the leptonic di-Higgs production cross-section, which can be studied at several (hypothetical) future high-energy colliders.  While for the case of the FCC-ee (even for the $t \bar t$ threshold run) and of the LCF at $\sqrt s=1$~TeV the NP contributions are rather small, for a future
muon collider, or a plasma wave acceleration driven electron positron collider, the cross-section can be significantly enhanced. 
In fact, and for a PWA $e^+e^-$ machine operating at $\sqrt{s}=10$~TeV, the enhancements of the production cross-section exceed the SM contribution by a factor of 60. 
Notice that such a striking enhancement occurs for moderately heavy HNLs (with masses around 5-10~TeV), 
while complying with all available bounds on EWPO and flavour-conserving Higgs and $Z$-boson decays (we recall that we explored regimes for which contributions to cLFV observables were deliberately set to be negligible). 

It is important to highlight that in the absence of contributions to low-energy cLFV observables (which are expected to be powerful probes of HNL extensions of the SM featuring generic flavour violating couplings), and for moderately heavy HNLs  - which may be too heavy to be directly produced at future high-energy colliders, the study of 
$\sigma(\ell^+ \ell^- \to hh)$ emerges as a complementary probe to these NP models. 
While synergetic studies of cLFV signatures and new contributions to EW precision observables are frequently elected as privileged means to test low-scale seesaw realisations (among them, the ISS(3,3)), it is clear that di-Higgs production at very-high energy lepton colliders offers a means to go beyond these tests, being sensitive to regimes with little associated contributions to cLFV or to EWPO. 
Interestingly, and in the years leading to very energetic electron and/or muon colliders, the expected FCC-ee reduction of uncertainties in EWPO might already shed light on the presence of such HNL states (due to the non-negligible contributions to the invisible $Z$ width and $T$ parameter).

\bigskip
HNL extensions of the Standard model are well-motivated NP constructions, with potential implications to numerous low- and high-energy observables. 
In view of their diversity, and of the variety of the associated signatures, the study of these extensions calls for thorough analysis of a wide array of observables; here, we have advocated that di-Higgs production at high-energy future facilities 
offers further possibilities to probe this class of models, even in limits in which they have little impact for cLFV and EW observables.

\section*{Acknowledgements}
JK is supported by the Slovenian Research Agency under the research core funding 
No. P1-0035 and in part by the research grants J1-3013 and N1-0253. 
EP acknowledges financial support from the Swiss
National Science Foundation (SNF) under contract 200020 204428, and is grateful for the hospitality of the Theory group of the Laboratoire de Physique de Clermont-Auvergne, where large parts of this work have been completed.
This project has received support from the IN2P3 (CNRS) Master Project, ``Hunting for Heavy Neutral Leptons'' (12-PH-0100).

\appendix

\section{The Inverse Seesaw} \label{app:model}
\subsection{Brief description of the ISS(3,3)}
In our study we consider a well-motivated SM extension featuring HNL: the Inverse Seesaw mechanism (ISS)~\cite{Schechter:1980gr, Gronau:1984ct, Mohapatra:1986bd}. In the ISS, two distinct species of sterile fermions $X$ and $\nu_R$ extend the original particle content of the SM, further leading to modified interactions, as given below
\begin{equation}\label{eq:ISS:lagrangian}
    \mathcal L_\text{ISS}\, =\, -Y^D_{ij} \,\overline{L_i^c}\,\widetilde H \,\nu_{Rj}^c - M_R^{ij}\, \overline{\nu_{Ri}}\, X_j - \frac{1}{2}\mu_R^{ij}\, \overline{\nu_{Ri}^c}\,\nu_{Rj} - \frac{1}{2} \mu_X^{ij}\, \overline{X_i^c}\, X_j + \text{H.c.}\,.
\end{equation}
Working the basis spanned by the interaction eigenstates, 
$(\nu_L, \nu_R^c, X)$, one is led to the following mass matrix for the neutral fermions
\begin{equation}
    M_\text{ISS} 
    = \begin{pmatrix}
    \mathbb{0} & m_D & \mathbb{0}\\
    m_D^T & \mu_R & M_R\\
    \mathbb{0} & M_R^T & \mu_X
    \end{pmatrix} 
    \,.
    \label{eqn:mass_matrix}
\end{equation}
In the above, $v$ denotes the electroweak vacuum expectation value, and one defines $m_D = v\,Y_D/\sqrt{2}$.
Working as usual in the well-motivated limit\footnote{Notice that working in the regime of small $\mu_{X,R}$ is natural in the sense of 't Hooft~\cite{tHooft:1980xss,Hettmansperger:2011bt}, since 
upon taking $\mu_{X,R}\to 0$, lepton number conservation is restored as a symmetry of the Lagrangian in Eq.~\eqref{eq:ISS:lagrangian}.}
$\mu_{X,R}\ll m_D \ll M_R$, $M_\text{ISS}$ can be block-diagonalised perturbatively; at leading order one finds\footnote{We neglect contributions from the Majorana mass term $\mu_R$ (whose effects only manifest in higher orders in the seesaw expansion and in loop corrections to the neutrino mass matrix~\cite{Dev:2012sg}), setting in to zero in our numerical computations.} that the light (mostly active) neutrino masses are given by  
\begin{eqnarray}
    m_\nu \simeq m_D\, (M_R^{-1})^T\, \mu_X\, M_R^{-1}\, m_D^T \equiv U_\text{PMNS}^\ast\, m_\nu^\text{diag}\, U_\text{PMNS}^\dagger\,.
    \label{eq:ISS:lightmasses}
\end{eqnarray}
Due to the smallness of lepton number violation (encoded in $\mu_X$),  
the heavy states combine to form approximately degenerate pseudo-Dirac pairs, the splittings  proportional to $\mu_X$.
As mentioned in the Introduction, we will consider 
a symmetric ISS realisation via the inclusion of $n_R = n_X = 3$ generations of heavy sterile fermions, the ISS(3,3)\footnote{Notice that this is not the minimal ISS realisation, as discussed in~\cite{Abada:2014vea}.}.
Notice that in the presence of non-negligible couplings between active and sterile neutral fermions, a new enlarged ($9\times9$) leptonic mixing matrix generalises the PMNS one. The latter is no longer unitary, and its deviations from unitarity will lead to the modification of charged and neutral lepton currents, which will in turn open the door to potentially large contributions to numerous observables. It proves convenient to introduce the so-called ``$\eta$-matrix''~\cite{FernandezMartinez:2007ms}, which encodes 
the deviations from unitarity of the $3\times 3$ would-be PMNS block of the enlarged unitary leptonic mixing matrix,
\begin{equation}
\label{eq:defPMNSeta}
U_\text{PMNS} \, \to \, \tilde U_\text{PMNS} \, = \,(\mathbb{1} - \eta)\, 
U_\text{PMNS}\,.
\end{equation}

With the goal of allowing a better connection between the fundamental parameters in the ISS Lagrangian and the (low-energy) data at the source of their most stringent constraints, new parametrisations of the Yukawa couplings were proposed in~\cite{Kriewald:2024rlg}: here we will consider the one relying on a singular value decomposition (SVD) of $Y_D$, which leads to parametrising the Yukawa couplings as follows
\begin{equation}
    Y_D^\text{SVD} = \frac{1}{v}\mathcal V_1\,\mathrm{diag}(y_1, y_2, y_3)\,\mathcal V_2^\dagger M_R^T\,.
    \label{eqn:SVD}
\end{equation}
In the above, and without loss of generality, one can take $y_i>0$. The unitary matrices $\mathcal V_{1,2}$
can each be parametrised as
\begin{equation}\label{eq:V12par}
    \mathcal V_{1,2} \,= \,O_{23} \,O_{13} \,O_{12}\,\mathrm{diag}(e^{i\varphi_1}, e^{i\varphi_2}, e^{i\varphi_3})\,,
\end{equation}
in which $\varphi_i$ denote 3 ``Majorana-like'' phases, and the rotation matrices $O_{ij}$ can be as usual cast in terms of 3 Euler angles and 3 ``Dirac-like'' phases $\delta_{ij}$; for example $O_{23}$ can be given as
\begin{equation}
    O_{23} = \begin{pmatrix}
        1 &0&0\\
        0& \cos\theta_{23} &\sin\theta_{23}e^{i \delta_{23}} \\
        0 & -\sin\theta_{23}e^{-i \delta_{23}}&\cos\theta_{23}
    \end{pmatrix} \,.
\end{equation}
A direct connection with the $\eta$ quantities (and hence with low-energy data) can be easily established,
\begin{equation}
    \eta \,= \frac{1}{2} \mathcal V_1^\ast\, \mathrm{diag}(y_1^2,y_2^2,y_3^2) \mathcal V_1^T\,.
\end{equation}
The quantities $y_i$ can be readily obtained from an eigenvalue decomposition of $\eta$.
Notice that 
for a diagonal (and thus flavour-conserving) $\eta$ one has $y_i = \sqrt{2\eta_{ii}}$. 
This is a particularly simple form we have considered throughout our work.

\bigskip
Let us also notice that for the purpose of the processes we discuss, we list below the relevant terms in the interaction Lagrangian 
(for a complete set, see e.g.~\cite{Abada:2021zcm}) 
\begin{align}
& \mathcal{L}_{W^\pm}\, =\, -\frac{g_w}{\sqrt{2}} \, W^-_\mu \,
\sum_{\alpha=1}^{3} \sum_{j=1}^{3 + n_S} \mathcal{U}_{\alpha j} \bar \ell_\alpha 
\gamma^\mu P_L \nu_j \, + \, \text{H.c.}\,, \nonumber \\
& \mathcal{L}_{Z^0}^{\nu}\, = \,-\frac{g_w}{4 \cos \theta_w} \, Z_\mu \,
\sum_{i,j=1}^{3 + n_S} \bar \nu_i \gamma ^\mu \left(
P_L {\mathcal C}_{ij} - P_R {\mathcal C}_{ij}^* \right) \nu_j\,, \nonumber \\
& \mathcal{L}_{H^0}\, = \, -\frac{g_w}{4 M_W} \, H  \,
\sum_{i\ne j= 1}^{3 + n_S}    \bar \nu_i\,\left[{\mathcal C}_{ij}\,\left(
P_L m_i + P_R m_j \right) +{\mathcal C}_{ij}^\ast\left(
P_R m_i + P_L m_j \right) \right] \nu_j\ , \nonumber \\
& \mathcal{L}_{G^0}\, =\,\frac{i g_w}{4 M_W} \, G^0 \,
\sum_{i,j=1}^{3 + n_S}  \bar \nu_i \left[ {\mathcal C}_{ij}
\left(P_R m_j  - P_L m_i  \right) + {\mathcal C}_{ij}^\ast
\left(P_R m_i  - P_L m_j  \right)\right] \nu_j\,, \nonumber  \\
& \mathcal{L}_{G^\pm}\, =\, -\frac{g_w}{\sqrt{2} M_W} \, G^- \,
\sum_{\alpha=1}^{3}\sum_{j=1}^{3 + n_S} \mathcal{U}_{\alpha j}
\bar \ell_\alpha\left(
m_\alpha P_L - m_j P_R \right) \nu_j\, + \, \text{H.c.}\,, 
\label{eqn:lag}
\end{align}
in which $\alpha, \rho = 1, \dots, 3$ denote the flavour of the charged leptons, while $i, j = 1, \dots, 3+n_S$ correspond to the physical (massive) 
neutrino states; $P_{L,R} = (1 \mp \gamma_5)/2$, $g_w$ is the weak coupling constant, and $\cos^2 \theta_w =  M_W^2 /M_Z^2$. 
Moreover, the coefficients ${\mathcal C}_{ij} $ (already defined in the main text, cf.~Eq.~(\ref{eq:Cij})) are given by
\begin{equation}
    {\mathcal C}_{ij} = \sum_{\rho = 1}^3
  \mathcal{U}_{i\rho}^\dagger \,\mathcal{U}_{\rho j}^{\phantom{\dagger}}\:. \nonumber
\label{eq:cij}\end{equation}
\subsection{Phenomenological constraints on SM extensions via HNL}
\label{app:constraints}

As extensively discussed, low-scale seesaw mechanisms as the ISS(3,3) can be at the source of important contributions to many observables, including cLFV transitions and decays, EW precision observables, quantities sensitive to the violation of lepton flavour universality, among many others. For generic extensions of the SM via $n_S$ Majorana sterile fermions, the expressions for the cLFV observables (form-factors and loop functions) can be found in~\cite{Ilakovac:1994kj,Alonso:2012ji,Abada:2018nio,2207.10109,Riemann:1982rq,Illana:1999ww,Mann:1983dv,Illana:2000ic,Ma:1979px,Gronau:1984ct,Deppisch:2004fa,Deppisch:2005zm,Dinh:2012bp,Abada:2014kba,Abada:2015oba,Abada:2015zea,Abada:2016vzu,Arganda:2014dta}.
For the ISS(3,3) cLFV $Z$-boson and Higgs decays have been studied in~\cite{1405.4300,1607.05257,2207.10109}; the contributions to numerous $Z$-pole observables (including EW precision observables) were presented in~\cite{2307.02558}.
In Table~\ref{tab:obs:EW-LFUV}, we collect  current experimental measurements and SM predictions for several lepton flavour universality violating (LFUV) and EW observables, which are of relevance for our discussion.
 \renewcommand{\arraystretch}{1.3}
\begin{table}[h!]
    \centering
    \hspace*{-2mm}{\small\begin{tabular}{|c|c|c|}
    \hline
    Observable & Exp. Measurement & SM prediction  \\
    \hline\hline
    $R_{\mu e}(Z\to\ell\ell)$ & $1.0001\pm 0.0024$ (PDG~\cite{ParticleDataGroup:2024cfk}) & $1.0$~\cite{Freitas:2014hra}\\
    $R_{\tau e}(Z\to\ell\ell)$ & $1.0020\pm0.0032$ (PDG~\cite{ParticleDataGroup:2024cfk}) & $0.9977$~\cite{Freitas:2014hra}\\
    $R_{\tau \mu}(Z\to\ell\ell)$ & $1.0010\pm 0.0026$ (PDG~\cite{ParticleDataGroup:2024cfk}) & $0.9977$~\cite{Freitas:2014hra}\\
    \hline
    $\Gamma(Z\to e^+e^-)$ & $83.91\pm0.12\:\mathrm{MeV}$ (LEP~\cite{ALEPH:2005ab}) & $83.965\pm0.016\:\mathrm{MeV}$~\cite{Freitas:2014hra}\\
    $\Gamma(Z\to \mu^+\mu^-)$ & $83.99\pm0.18\:\mathrm{MeV}$ (LEP~\cite{ALEPH:2005ab}) & $83.965\pm0.016\:\mathrm{MeV}$~\cite{Freitas:2014hra}\\
    $\Gamma(Z\to \tau^+\tau^-)$ & $84.08\pm0.22\:\mathrm{MeV}$ (LEP~\cite{ALEPH:2005ab}) & $83.775\pm0.016\:\mathrm{MeV}$~\cite{Freitas:2014hra}\\
    \hline
    $\Gamma(Z\to\mathrm{inv.})$ & $499.3 \pm 1.5\:\mathrm{MeV}$ (PDG~\cite{ParticleDataGroup:2024cfk})& $501.45\pm 0.05\:\mathrm{MeV}$~\cite{Freitas:2014hra}\\
    \hline
    \end{tabular}}
    \caption{Experimental measurements and SM predictions for several LFUV and EW observables discussed in the phenomenological analysis. All uncertainties are given at 68\% C.L., while for the SM predictions of the universality ratios, the parametric uncertainties are negligible.}
    \label{tab:obs:EW-LFUV}
\end{table}
\renewcommand{\arraystretch}{1.}

Upon taking into consideration a vast array of low-energy processes (which are sensitive to new sources of lepton flavour universality violation), one is led to bounds on the diagonal entries of the $\eta$-matrix. Following the  analysis of~\cite{Blennow:2023mqx} one has (at $95\%$~C.L.):
\begin{eqnarray}
    \eta_{ee}&\lesssim& 1.4\times 10^{-3}\,,\nonumber\\
    \eta_{\mu\mu} &\lesssim& 1.4\times 10^{-4}\,,\nonumber\\
    \eta_{\tau\tau} &\lesssim& 8.9\times 10^{-4}\,.\label{eqn:etaconstraint}
\end{eqnarray}
Since we consider strictly flavour conserving Yukawa structures, any flavour violation is proportional to the entries of the small Majorana mass term $\mu_X$, therefore leading to negligible contributions to charged lepton flavour violation observables.

\section{Detailed expressions for the di-Higgs production cross-sections }\label{app:analytical} 
In this appendix we collect the box form-factors relevant for the computation of the SM and NP contributions to $\ell^+\ell^-\to hh$.
Beginning with the SM contributions from the diagrams in Fig.~\ref{fig:SM:eeHH}, we evaluate the box diagrams in the chiral limit of vanishing lepton masses $m_\ell\to 0$.

The contributions from the boxes with three $Z$-bosons are, in the limit $m_\ell\to 0$, 
\begin{eqnarray}
    F_{L,t}^Z &=& \dfrac{g^4}{16 \pi^2}\dfrac{(M_Z^2-2M_W^2)^2}{16 M_W^4} \bigg[ 
     -16 D_{00} - 6 D_{002} + (D_{123} + 4 D_{13}) s - (4 D_{22}  + D_{222}) t 
     \nonumber\\
     &&+ (4 D_{1} + 
 4 D_{12}+ D_{122} + D_{223} + 4 D_{23}+ 
 4 D_{3})(m_h^2 - t)  - 
 4 D_{2} (2 M_Z^2 + t)
\bigg]\,,
\end{eqnarray}
with the Passarino-Veltman functions $D_X=D_X(m_h^2, m_h^2, 0, t, s, 0, M_Z^2, M_Z^2, M_Z^2)$, in the convention of LoopTools~\cite{Hahn:1998yk}. 
The right-handed form-factor is given by 
\begin{eqnarray}
    F_{R,t}^Z =\dfrac{4(M_W^2-M_Z^2)^2}{(M_Z^2-2M_W^2)^2} F_{L,t}^Z\,,
\end{eqnarray}
and the form-factors of the associated diagrams with crossed legs read
\begin{eqnarray}
    F_{L/R,u}^Z = - F_{L/R,t}^Z \,(t\leftrightarrow u)\,.
\end{eqnarray}

The contributions arising from the boxes with three $W$-bosons, for $m_\ell\to 0$, are related to the $Z$ diagrams, differing by the gauge boson masses and a global factor accounting for the different couplings.    
One thus finds the correspondences:
\begin{eqnarray}
    F_{L,t/u}^W &=& \dfrac{2 M_W^4}{(M_Z^2-2M_W^2)} [F_{L,t/u}^Z\,(M_Z \to M_W)]\,.
\end{eqnarray}

Concerning the NP contributions, 
we further categorise them by the multiplicity of the HNL propagating virtually in the loops; accordingly, we separate them into $3N$, $2N$ and $1N$ contributions.
The form-factor $F_{L,t}^{3N}$ corresponds to the box diagram with three HNLs in the loop (i.e. first box of Fig.~\ref{fig:eehh}), and the associated $F_{L,u}^{3N}$ corresponds to the same diagram with the crossed legs.
Similarly, $F_{L,t(u)}^{2N} $ correspond to the box diagrams with two HNLs and two $W$-bosons in the loop (with crossed legs). Finally $F_{L,t(u)}^{N} = F_{L,t(u)}^{N,\mathrm{BSM}}+F_{L,t(u)}^{N,\mathrm{NP}}$  are the form-factors of the boxes with three $W$-bosons, where $F_{L,t(u)}^{N,\mathrm{BSM}}$ corresponds to ``SM''-like boxes with only HNLs (no light neutrinos to avoid double counting), while $F_{L,t(u)}^{N,\mathrm{NP}}$ correspond to the boxes with three $W$-bosons in the longitudinal limit.
The latter vanish in the SM due to the smallness of electron and neutrino masses (i.e. the ones with one or more Goldstone bosons, except the one attached to the 2 Higgses).
Notice that all form-factors exhibit a manifest crossing symmetry $ F_{L,u}^{xN} = - F_{L,t}^{xN} (t\leftrightarrow u)$, which serves as an important and non-trivial consistency check of the computation.
\begin{eqnarray}
    F_{L,t}^{3N} &=& \dfrac{1}{16\pi^2}\dfrac{4 g^4}{32 m_W^4} \sum_{i,j,k} \mathcal{U}_{\ell j}\mathcal{U}_{\ell k}^*\bigg\{ 2 (D_{1} +D_{3})(C_{ji} m_i+C_{ji}^* m_j) (C_{ik} m_i+C_{ik}^* m_k) s M_W^2
    \nonumber\\
    &&+\left[-2 D_{00} +6 D_{002} -D_{123} s +D_{13} s -D_{122}  \left(m_h^2-t\right)+D_{223}  \left(t-m_h^2\right) +D_{22} t +D_{222} t\right] 
    \nonumber\\
    &&  \hspace{2mm}\times\big[C_{ik}^* C_{ji}^* m_j m_k \left(m_i^2+2 M_W^2\right)+C_{ik}^* C_{ji} m_i m_k \left(m_j^2+2 M_W^2\right)\nonumber\\
    &&  \hspace{2mm}+C_{ik} C_{ji}^* m_i m_j \left(m_k^2+2 M_W^2\right)+C_{ik} C_{ji} \left(m_j^2 m_k^2+2 m_i^2 M_W^2\right)\big]
    \nonumber\\
    &&+D_{0} \big[C_{ik}^* m_k \left(C_{ji}^* m_k^2 m_j^3+C_{ji} m_i \left(m_k^2+2 M_W^2\right) m_j^2+2 C_{ji}^* M_W^2 \left(m_i^2+s\right) m_j+2 C_{ji} m_i M_W^2 s\right)
    \nonumber\\
    &&  \hspace{2mm}+C_{ik} \left(C_{ji}^* m_i m_k^2 m_j^3+2 C_{ji} m_k^2 M_W^2 m_j^2+2 C_{ji}^* m_i M_W^2 \left(m_k^2+s\right) m_j+C_{ji} m_i^2 \left(m_j^2 m_k^2+2 M_W^2 s\right)\right)\big]
    \nonumber\\
    &&+D_{2} \big[C_{ik}^* C_{ji} m_i m_k \left(\left(m_j^2+2 M_W^2\right) m_i^2+2 m_j^2 \left(m_k^2+2 M_W^2\right)+2 M_W^2 s\right)
    \nonumber\\
    &&  \hspace{2mm}+C_{ik}^* C_{ji}^* m_j m_k \left(\left(m_j^2+m_k^2+6 M_W^2\right) m_i^2+m_j^2 m_k^2+2 M_W^2 s\right)
    \nonumber\\
    &&  \hspace{2mm}+C_{ik} C_{ji}^* m_i m_j \left(\left(m_k^2+2 M_W^2\right) m_i^2+2 m_j^2 m_k^2+2 M_W^2 \left(2 m_k^2+s\right)\right)\nonumber\\
    &&  \hspace{2mm}+C_{ik} C_{ji} \left(\left(\left(3 m_k^2+2 M_W^2\right) m_j^2+2 M_W^2 \left(m_k^2+s\right)\right) m_i^2+2 m_j^2 m_k^2 M_W^2\right)\big]\bigg\}\,,
\end{eqnarray}
where the Passarino-Veltman functions are $D_{X} = D_{X}(0,m_h^2,m_h^2,0,t,s,M_W^2,m_j^2,m_i^2,m_k^2)$ and with $F_{L,u}^{3N} = - F_{L,t}^{3N}\,(t\leftrightarrow u)$.
The ``$2N$-diagrams'' result in
\begin{eqnarray}
    F_{L,t}^{2N} &=& \dfrac{1}{16\pi^2}\dfrac{2 g^4}{32 m_W^4} \sum_{i,j} \mathcal{U}_{\ell j}\mathcal{U}_{\ell i}^*\bigg\{ -12 D_{00} M_W^2 \left(-C_{ji} m_i^2+2 C_{ji} m_j^2+C_{ji}^* m_i m_j\right)
   \nonumber\\
    && -2 D_{0} m_i \left(C_{ji} m_i \left(m_h^2 m_j^2+m_j^2 M_W^2+4 M_W^4\right)+C_{ji}^* m_j \left(m_h^2 m_i^2+M_W^2 \left(-m_i^2+2 m_j^2+4 M_W^2\right)\right)\right)
    \nonumber\\
    && - \big[12 D_{001}+ 12 D_{002}+ 2 D_{111} m_h^2 +2 D_{112}  \left(2 m_h^2+t\right) + 2 D_{113} \left(s+t-m_h^2\right) + 2 D_{122}  \left(m_h^2+2 t\right) 
    \nonumber\\
    && \hspace{2mm}+2 D_{123}  \left(s+2t -2 m_h^2\right) + 2 D_{223}  \left(t-m_h^2\right) +2 D_{222}  t  \big] M_W^2 \left(C_{ji} \left(m_i^2+m_j^2\right)+2 C_{ji}^* m_i m_j\right)
    \nonumber\\
    &&-2 D_{12} M_W^2 \left(C_{ji} m_h^2 \left(m_i^2+3 m_j^2\right)-C_{ji} t \left(m_i^2-3 m_j^2\right)+2 C_{ji}^* m_i m_j \left(2 m_h^2+t\right)\right)
   \nonumber\\
   &&-2 D_{1} \big[C_{ji} \left(m_h^2 \left(m_i^2 \left(2 m_j^2-M_W^2\right)+2 m_j^2 M_W^2\right)+2 m_i^2 M_W^2 \left(m_j^2+2 M_W^2\right)+4 m_j^2 M_W^4\right)\nonumber\\
    && \hspace{2mm}+C_{ji}^* m_i m_j \left(m_h^2 \left(m_i^2+m_j^2+M_W^2\right)+M_W^2 \left(m_i^2+m_j^2+8 M_W^2\right)\right)\big]
   \nonumber\\
   &&-6 D_{11} m_h^2 m_j M_W^2 (C_{ji} m_j+C_{ji}^* m_i)
\nonumber\\
   &&+2 D_{13} M_W^2 \left(s+t-m_h^2\right) \left(C_{ji} \left(m_i^2-3 m_j^2\right)-2 C_{ji}^* m_i m_j\right)
   \nonumber\\
   &&-2 D_{2} \big[C_{ji} \left(m_h^2 m_i^2 \left(2 m_j^2-M_W^2\right)+2 M_W^2 \left(m_i^2 \left(m_j^2+2 M_W^2+s\right)+m_j^2 \left(2 M_W^2+t\right)\right)\right)
   \nonumber\\
    && \hspace{2mm}+C_{ji}^* m_i m_j \left(m_h^2 \left(m_i^2+m_j^2-M_W^2\right)+M_W^2 \left(m_i^2+m_j^2+2 \left(4 M_W^2+s+t\right)\right)\right)\big]
   \nonumber\\
   &&-2 D_{22} M_W^2 \left(C_{ji} \left(m_h^2 m_i^2-m_i^2 t+3 m_j^2 t\right)+C_{ji}^* m_i m_j \left(m_h^2+2 t\right)\right)
   \nonumber\\
   &&+2 D_{23} M_W^2 \left(C_{ji} \left(m_h^2 \left(m_i^2-3 m_j^2\right)-m_i^2 (s-t+2m_h^2)+3 m_j^2 (2m_h^2 - t)\right)+C_{ji}^* m_i m_j \left(2 m_h^2- s - 2 t\right)\right)
   \nonumber\\
   &&-4 D_{3} m_j M_W^2 \left(t-m_h^2\right) (C_{ji} m_j+C_{ji}^* m_i)\bigg\}\,,
\end{eqnarray}
with $D_{X} = D_{X}(m_h^2,0,m_h^2,0,t,u,m_i^2,m_j^2,M_W^2,M_W^2)$ and $F_{L,u}^{2N}=-F_{L,t}^{2N} \,(t \leftrightarrow u)$.
Furthermore, the $1N$-diagram (partially) yields
\begin{eqnarray}
    F_{L,t}^{N,\mathrm{BSM}} &=& \dfrac{1}{16\pi^2}\dfrac{ g^4}{16} \sum_{i>3} \mathcal{U}_{\ell i}\mathcal{U}_{\ell i}^*\bigg\{ -32 D_{00} -2  \left(6 D_{002}-s (D_{123}+4 D_{13})+4 D_{2} \left(2 M_W^2+t\right)\right)\nonumber\\
        &&+8 D_{1}  \left(m_h^2-t\right)+8 D_{12}  \left(m_h^2-t\right)+2 D_{122}  \left(m_h^2-t\right)-8 D_{22} t-2 D_{222}  t
        \nonumber\\
        &&-2 D_{223} \left(t-m_h^2\right)-8 D_{23}  \left(t-m_h^2\right)-8 D_{3}  \left(t-m_h^2\right)
    \bigg\}\,,
\end{eqnarray}
where $D_{X} = D_{X}(0,m_h^2,m_h^2,0,t,s,m_i^2,M_W^2,M_W^2,M_W^2)$ and $F_{L,u}^{N,\mathrm{BSM}} = - F_{L,t}^{N,\mathrm{BSM}} \,(t\leftrightarrow u)$.
Notice that the sum excludes the light neutrino contributions to avoid double counting of the SM amplitudes.
Finally, the longitudinally enhanced Goldstone contributions (absent in the SM) are given by
\begin{eqnarray}
    F_{L,t}^{N,\mathrm{NP}} &=& \dfrac{1}{16 \pi^2}\dfrac{g^4}{16 M_W^2} \sum_i \mathcal{U}_{\ell i}\mathcal{U}_{\ell i}^* m_i^2 \bigg\{ 8 D_{0} \left( M_W^2- m_h^2\right)-4 D_{00}+12 D_{002}+2(D_{12}+D_{23}) (m_h^2-s-t)
    \nonumber\\
    &&+2(D_{122}+ D_{223}) \left( t- m_h^2\right)-2 D_{123} s-4 D_{22} m_h^2 +2 D_{222} t
    \nonumber\\
    &&+D_{2} \left(-\frac{2 m_h^4}{M_W^2}
    -8 M_W^2
    -4 s -2t\right)
\bigg\}\,,
\label{eqn:goldt}
\end{eqnarray}
with $D_{X} = D_{X}(0,m_h^2,m_h^2,0,t,s,m_i^2,M_W^2,M_W^2,M_W^2)$ and $F_{L,u}^{N,\mathrm{NP}} = - F_{L,t}^{N,\mathrm{NP}} \,(t\leftrightarrow u)$.
Notice that these are relatively enhanced for $m_i\gg M_W$, due to the global factor $m_i^2/M_W^2$ which alleviates the $\mathcal U^2 \sim v^2 Y_D^2/m_i^2$ suppression.

\end{document}